\documentclass[sigconf]{acmart}
\AtBeginDocument{%
  }

\setcopyright{acmlicensed}
\copyrightyear{2018}
\acmYear{2018}
\acmDOI{XXXXXXX.XXXXXXX}
\acmConference[Conference acronym 'XX]{Make sure to enter the correct
  conference title from your rights confirmation emai}{June 03--05,
  2018}{Woodstock, NY}
\acmISBN{978-1-4503-XXXX-X/18/06}

\usepackage{multirow}
\usepackage{threeparttable}
\usepackage{enumitem}
\usepackage{newpxtext}

\makeatletter
\newcommand{\pub}[1]{\color{gray}{\tiny{[{#1}]}}}
\newcommand{\ie}{\emph{i.e.}\@ifnextchar.{\!\@gobble}{}}
\newcommand{\eg}{\emph{e.g.}\@ifnextchar.{\!\@gobble}{}}
\newcommand{\etc}{etc\@ifnextchar.{}{.\@}}
\makeatother

\begin{document}

\title{Understanding Long Videos via \\LLM-Powered Entity Relation Graphs}

\author{Meng Chu}

\affiliation{%
  \institution{National University of Singapore}
  \country{Singapore}
}

\author{Yicong Li}
\affiliation{%
  \institution{National University of Singapore}
  \country{Singapore}
}

\author{Tat-Seng Chua}
\affiliation{%
  \institution{National University of Singapore}
  \country{Singapore}
}

\renewcommand{\shortauthors}{Meng et al.}

\begin{abstract}
The analysis of extended video content poses unique challenges in artificial intelligence, particularly when dealing with the complexity of tracking and understanding visual elements across time. Current methodologies that process video frames sequentially struggle to maintain coherent tracking of objects, especially when these objects temporarily vanish and later reappear in the footage. A critical limitation of these approaches is their inability to effectively identify crucial moments in the video, largely due to their limited grasp of temporal relationships.
To overcome these obstacles, we present GraphVideoAgent, a cutting-edge system that leverages the power of graph-based object tracking in conjunction with large language model capabilities. At its core, our framework employs a dynamic graph structure that maps and monitors the evolving relationships between visual entities throughout the video sequence. This innovative approach enables more nuanced understanding of how objects interact and transform over time, facilitating improved frame selection through comprehensive contextual awareness.
Our approach demonstrates remarkable effectiveness when tested against industry benchmarks. In evaluations on the EgoSchema dataset, GraphVideoAgent achieved a 2.2\% improvement over existing methods while requiring analysis of only 8.2 frames on average. Similarly, testing on the NExT-QA benchmark yielded a 2.0\% performance increase with an average frame requirement of 8.1. These results underscore the efficiency of our graph-guided methodology in enhancing both accuracy and computational performance in long-form video understanding tasks.
\end{abstract}



\received{20 February 2007}
\received[revised]{12 March 2009}
\received[accepted]{5 June 2009}


\maketitle

\section{Introduction}
\label{sec:intro}

\begin{figure}[t!]
\centering
\includegraphics[width=\linewidth]{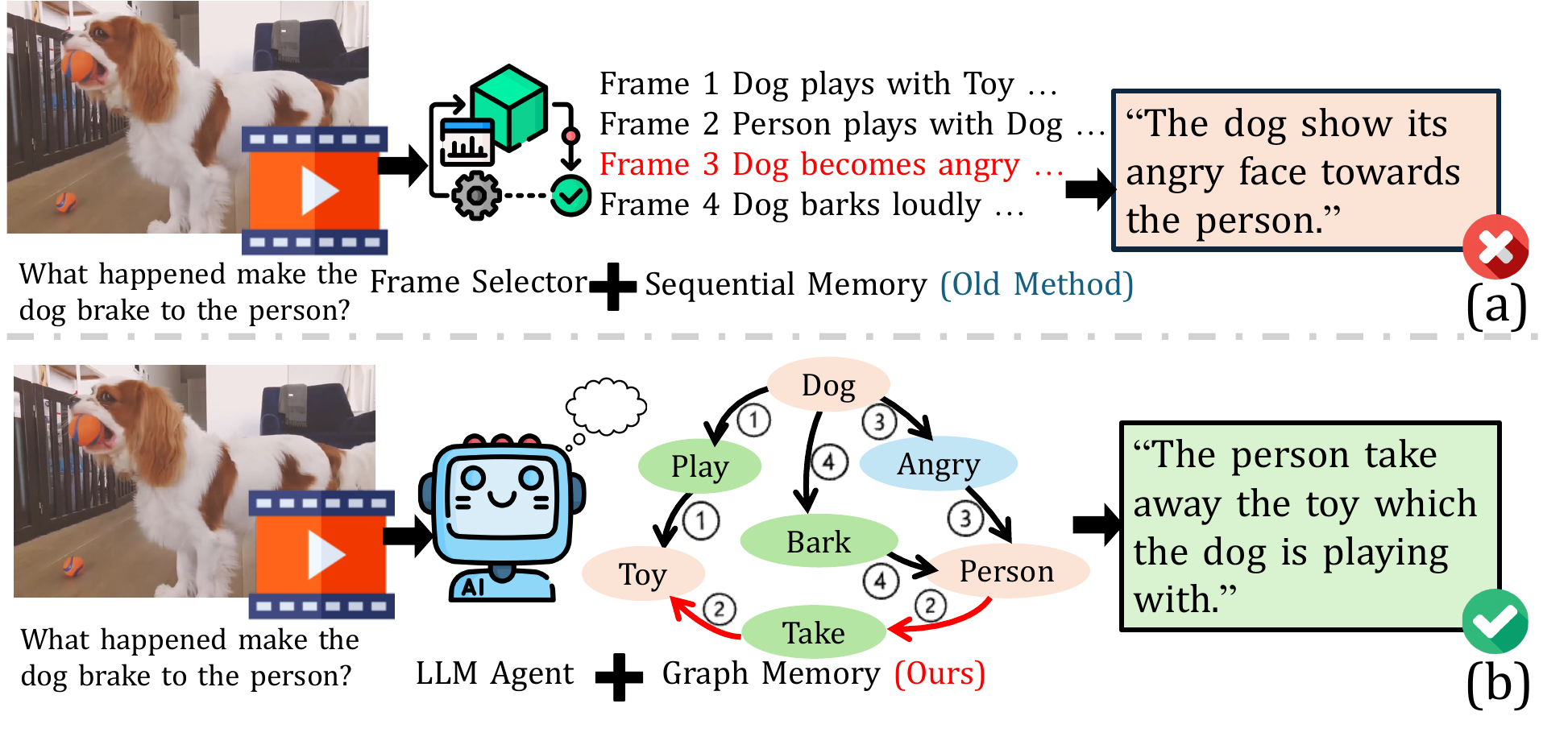} 
\caption[]{Paradigm comparison --- \protect\textbf{(a)} The traditional method employs a frame selector with sequential memory, which processes frames linearly and outputs ``The dog shows its angry face towards the person,'' missing the causal relationship. \protect\textbf{(b)} In contrast, our method combines an LLM Agent with Graph Memory, representing entities (\protect\textit{i.e.}, Dog, Toy, Person) and their interactions through a structured graph.}
\label{fig:architecture}
\end{figure}

Long-form video understanding (LVU) seeks to answer complex questions based on extensive video content, ranging from minutes to hours in duration. This task presents significant challenges due to the multimodal nature of the data and the vast length of the video \cite{song2024moviechat, weng2025longvlm}. Specifically, it requires capabilities in three critical areas: 1) structural comprehension of the video content, 2) substantial memory capacity for storing visual clues, and 3) advanced multimodal reasoning abilities to filter answer information from a large volume of visual data.

Recent advancements in LVU \cite{sun2022long, ma2023vista, wu2022memvit, song2024moviechat, jin2023chatunivi} have primarily focused on sequential frame processing with a frame selector, where models typically capture frame-wise visual features either independently \cite{liu2022ts2, ghodrati2021frameexit} or with limited temporal context \cite{li2023videochat, wang2023selective}, while maintaining a sequential memory bank to record these visual cues. However, long-form videos often involve multiple objects that interact dynamically across time---certain objects may become occluded or exit the camera's view, only to reappear later in video. Storing visual clues as a simple sequence struggles to capture the evolving relations among these visual entities, making sequential memory updates insufficient for adapting to the complexities of long videos. As illustrated in Figure \ref{fig:architecture}a, the sequential memory approach fails to properly identify the cause of the dog's aggressive behavior. This limitation stems from the sequential memory's inability to maintain and update the complex, interconnected relations between the dog, toy, and person as they interact over time.

In contrast, humans process long videos by naturally maintaining a mental graph of entities and their evolving relations across time. This cognitive process involves selective attention to focus on key moments and semantic tracking to maintain coherence, both of which are well-documented in memory research \cite{cohendet2019videomem, baddeley2020memory}. Unlike current video understanding models, which often lack these capabilities \cite{zhang2024flash}, humans continuously update mental models, track relations, and dynamically allocate attention based on context, enabling a more coherent and adaptive understanding of the video's narrative.

\begin{figure*}[t!]
\centering

\includegraphics[width=\linewidth]{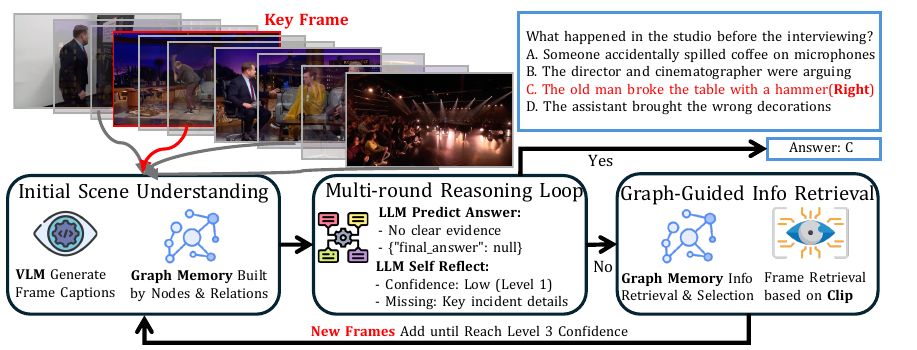} \\

\caption{The figure illustrates GraphVideoAgent's architecture, which consists of four main components: (1) an input module that performs uniform sampling from long videos, (2) a dynamic entity-relation graph that tracks entities and their temporal relations, (3) foundation model tools including CLIP, VLM, and frame retrieval for processing video content, and (4) an LLM agent responsible for frame selection, graph updates, and answer generation. These components work together to enable graph-enhanced video understanding capabilities.}
\label{fig:map}
\end{figure*}

Inspired by human cognitive processes, we propose Graph-VideoAgent, a novel LVU framework that 
integrates two key components: (1) an LLM agent that iteratively identifies and analyzes critical information via multi-round reasoning and self-reflection, and (2) a dynamic graph memory that explicitly tracks temporal and semantic relations among visual entities.
Unlike existing sequential frame processing methods, which struggle with frame selection and temporal coherence, our graph-guided frame selection accurately identifies key frames by tracking entity relations across time, requiring fewer frames of answer clues, and thus improving efficiency. Additionally, in contrast to prior works that rely on static queries \cite{wang2025videoagent}, our dynamic graph structure enables advanced query refinement by considering the evolving nature of relations and contexts within the video. As shown in Figure \ref{fig:architecture}b, our LLM agent selects key frames from the video while the graph memory helps track relations between objects. By maintaining a graph structure with nodes (\ie, Dog, Toy, Person) and temporal edges showing actions (\ie, Play, Take, Bark), our approach can easily identify that the person taking the toy led to the dog becoming angry. This simple but effective graph representation helps establish clear causal relations between events in the video. Figure \ref{fig:map} gives the overview of the architecture.

Our contributions are summarized as follows:
\begin{itemize}[leftmargin=*]
    \item We analyze the challenge of LVU under current sequential memory-based design. Inspired by human cognitive processes, we highlight the importance of modulating the evolving relations among visual entities as a key component of LVU.

    \item We propose GraphVideoAgent, an agent-based LVU framework that explicitly modulates a dynamic entity relation graph, integrating large language model (LLM)-based reasoning with graph-structured entity tracking. This enables more structured processing of video content over sequential frame-based models.

    \item Through extensive experiments on two LVU benchmarks, our model achieves state-of-the-art performance (EgoSchema \cite{mangalam2024egoschema} +2.2\% and NExT-QA \cite{xiao2021next} +2.0\%). Furthermore, it demonstrates remarkable efficiency, utilizing only 8.2 and 8.1 frames on average. 

\end{itemize}

\section{Related Works}
\subsection{Long-form Video Understanding} 
Long-form video understanding poses significant challenges due to its computational demands and the complexity of temporal relations. Several approaches have been proposed to address these challenges. End-to-end models~\cite{sun2022long,ma2023vista,wu2022memvit} attempt to process entire videos through transformer architectures but often struggle with memory constraints. Compression-based methods~\cite{nguyen2022s4nd,islam2022long} reduce computational demands but risk losing temporal information. Frame selection approaches~\cite{yu2023self,xu2023retrievalbased,lei2021less} improve efficiency by identifying key frames but typically treat them independently, losing important temporal relations. Recent works like VideoAgent~\cite{zhang2023simple} introduce iterative frame selection guided by LLMs, and VideoAgent~\cite{wang2025videoagent} augments this with a memory mechanism, but both lack explicit modelling of entity relations across time.

\subsection{Graph-based Visual Understanding}
Graph structures have demonstrated effectiveness in various visual understanding tasks, though primarily in static contexts. For static images, graphs have been used to model relations between objects~\cite{jing2020visual}, enhancing visual reasoning capabilities. Scene graph generation, which aims to parse visual scenes into structured representations of objects and their relations, has seen significant developments from foundational works like Neural Motifs~\cite{zellers2018neural} and LinkNet~\cite{woo2018linknet} to more recent advances in panoptic scene understanding~\cite{yang2022panoptic} and dynamic scene modeling~\cite{ost2021neural}. Recent work~\cite{gao2024graphdreamer}has even extended scene graphs to enable compositional 3D scene synthesis. In video understanding, some works have applied graph structures for action recognition in short clips~\cite{wang2021supervoxel,yang2020gives} and scene graph generation~\cite{hussein2019videograph,korbar2023text}. However, these approaches typically focus on frame-level or short-term relations rather than tracking entities and their evolving relations across extended temporal sequences. Memory-based approaches~\cite{wang2023lifelongmemory} have explored structured representations for videos but usually employ flat memory structures that don't capture complex entity relations.

\subsection{LLM Agents for Visual Understanding}
The emergence of large language models has sparked interest in using them as reasoning agents for visual understanding tasks. Recent works have demonstrated LLMs' potential as coordinators for vision-language models~\cite{suris2023vipergpt,zhang2023simple,chu20243d,chu2025towards} and explored memory augmentation for temporal understanding~\cite{wang2023lifelongmemory}. Recent benchmarks and frameworks like VisualAgentBench~\cite{liu2024visualagentbench} and VisualWebArena~\cite{koh2024visualwebarena} have advanced the evaluation of multimodal agents across diverse scenarios, while architectural innovations like CogAgent~\cite{hong2024cogagent} have improved visual-textual understanding capabilities. Our work builds upon these advances by introducing a graph-based memory structure that explicitly tracks entities and their relations across time, while leveraging an LLM agent for reasoning and frame selection. This approach enables both efficient processing of long-form videos and sophisticated temporal reasoning about entity relations, addressing limitations of previous approaches that either lack structured representations or struggle with long-term temporal dependencies.

\begin{figure*}[t!]
\centering

\includegraphics[width=0.95\linewidth]{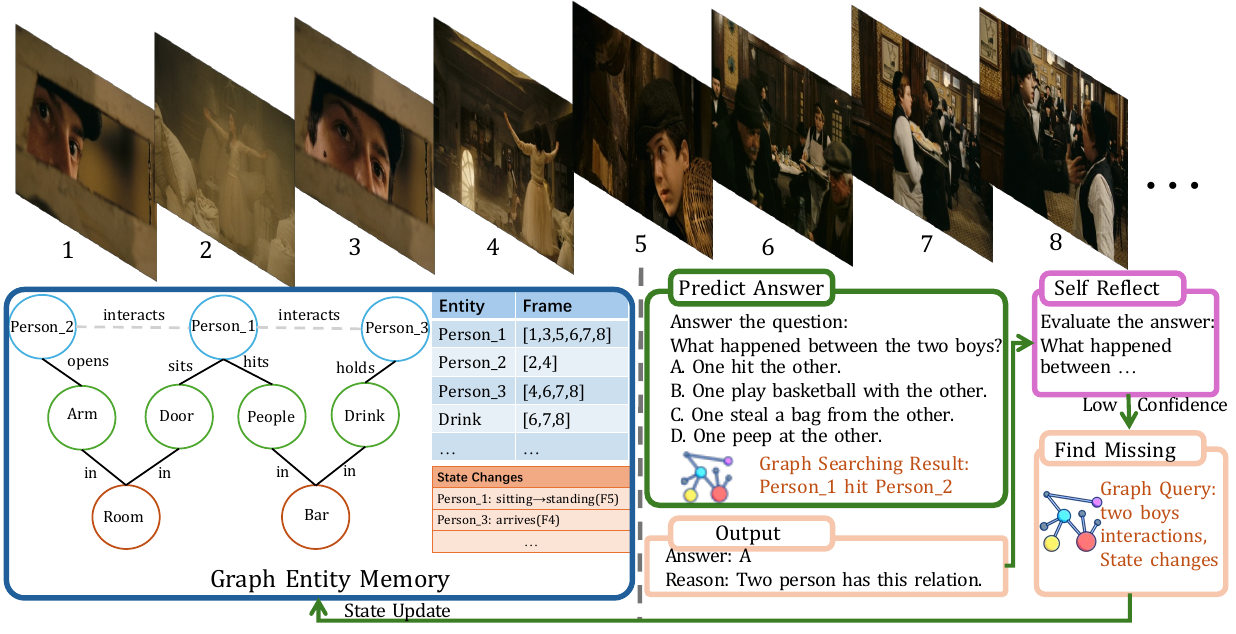} \\

\caption{GraphVideoAgent's video analysis process has multiple components: a sequence of 8 video frames showing interactions between people in an indoor setting (top), a Graph Entity Memory structure (bottom left) that maps relations between entities (people, objects, and actions) and tracks their appearances across frames, and a reasoning process (bottom right) that uses this graph structure to answer questions about the video. The system includes a multiple-choice question interface, graph searching capabilities, and self-reflection mechanisms to evaluate answer confidence. The graph maintains entity relations, state changes, and temporal information to enable accurate video understanding and question answering.}
\label{fig:flow}
\end{figure*}

\section{Method}

\subsection{Overview}
GraphVideoAgent presents a novel approach showing in Figure \ref{fig:flow} to video question answering by integrating sophisticated entity tracking, temporal state modeling, and multi-relational graph representation with an iterative LLM-based reasoning framework. The system constructs a dynamic knowledge graph G = (V, E) that captures rich entity relations and temporal dynamics from video frame captions. Given an input video V and question q, our system iteratively builds and updates G while selecting relevant frames F = {f1, ..., fn} to generate the answer a. Unlike previous approaches that treat frames independently, our system employs a comprehensive strategy that iteratively refines frame selection based on entity states, relation patterns, and temporal coherence. The system's architecture integrates several key components: VLM (e.g., EVA-CLIP \cite{sun2024eva}) for robust visual feature extraction and similarity computation, NLP tools (e.g., spaCy \cite{spacy2020}) for sophisticated entity extraction and linguistic analysis, and LLM (e.g., GPT-4 \cite{openai2023gpt4}) for multi-round reasoning and answer generation. The LLM agent operates through an iterative process where it first establishes an initial state through uniform frame sampling and caption generation, then progressively refines its understanding through multiple rounds of analysis. In each round, the agent evaluates its confidence through self-reflection, determining whether to provide an answer or gather additional information through targeted frame retrieval. When additional information is needed, the agent leverages the graph structure to guide its search, focusing on relevant temporal segments and entity relations. The key innovations of our approach lie in two interconnected aspects: video knowledge graph which includes the enhanced multi-level entity-relation graph with the temporal state tracking and the LLM agent which could finish segment-level adaptive frame retrieval, and the multi-round LLM reasoning. This integrated approach enables the system to efficiently decompose complex temporal reasoning tasks while maintaining computational efficiency through selective processing and targeted information gathering. The combination of structured graph representation with iterative reasoning particularly enhances the system's ability to handle challenging scenarios where understanding long-term dependencies, causal relations, and subtle interactions between events is crucial.

\subsection{Video Knowledge Graph Mechanism}
The system constructs graphs using entity nodes with temporal features and three types of relation edges (spatial, interaction, and action). Through its state tracking mechanism and three-level architecture, it enables comprehensive modeling of both entity interactions and temporal patterns in videos.

\noindent\textbf{Enhanced Entity-relation Graph.} Our system constructs a sophisticated multi-relational graph structure $G = (V, E)$ that captures rich entity interactions and their temporal evolution. Each entity node $v_{i} \in V$ is derived through a comprehensive extraction process that combines named entity recognition with noun phrase chunking. The extracted entities are organized in a hierarchical type system encompassing categories such as Person, Location, Object, and Group. Each node maintains a rich information tuple $(F_{i}, x_{i}, c_{i}, s_{i})$, where $F_{i}$ tracks frame indices of entity appearances, $x_{i}$ stores VLM-derived visual features, $c_{i}$ contains caption descriptions, and $s_{i}$ records state change sequences.

The edge structure $E$ represents three distinct categories of relations extracted through linguistic analysis. Spatial relations are identified through prepositional phrases (e.g., "in", "on", "at"), capturing physical positioning and spatial context. Interaction relations are derived from specific verbs (e.g., "talk", "meet", "speak") that indicate direct engagement between entities. Action relations are extracted from dynamic verbs (e.g., "open", "close", "hold"), representing specific activities entities perform. Each edge $e_{ij} \in E$ is constructed through dependency parsing of captions, storing relation type, temporal information, and associated linguistic elements.

\noindent\textbf{Temporal State Tracking.}
The temporal dimension of our graph system is handled through a sophisticated state tracking mechanism that captures both entity evolution and relation dynamics. At its core, the system continuously monitors state-indicating verbs to detect and record entity state transitions, creating detailed state history sequences. This tracking extends beyond individual states to capture relation evolution patterns, monitoring how entity interactions persist and transform across video sequences.

Temporal coherence is maintained through a carefully designed computation:
\begin{equation}
T(e,f) = \alpha \cdot S(e,f) + (1-\alpha) \cdot R(e,f)
\end{equation}
where $S(e,f)$ quantifies entity state consistency and $R(e,f)$ measures relation persistence at frame $f$. This computation is complemented by an adaptive window mechanism that dynamically adjusts temporal context based on relation significance, enabling the system to maintain both local temporal consistency and global contextual understanding.

\noindent\textbf{Multi-level Graph Structure.} The graph implementation employs a multi-level architecture that operates across three distinct levels to capture the complex dynamics of video content. At the entity level, the structure maintains comprehensive entity profiles within the hierarchical typing system, integrating state histories with visual-semantic features. The relation level manages the intricate network of inter-entity connections, supporting multiple relation types while preserving temporal and spatial context.

At the global level, the system ensures cross-frame consistency through a dynamic update mechanism:
\begin{equation}
G_{t+1} = U(G_{t}, F_{t+1}, R_{t+1})
\end{equation}
where new frames $F_{t+1}$ and relations $R_{t+1}$ are seamlessly integrated into the existing structure. This multi-level design enables the system to capture both fine-grained entity interactions and broader temporal patterns, providing a comprehensive foundation for video understanding.

\subsection{LLM Agent}
The LLM agent in our system operates through a carefully designed iterative process that combines frame selection with multi-round reasoning. This design enables progressive information gathering while maintaining computational efficiency. The agent's operation consists of two main components: iterative frame selection and question answering process.

\noindent\textbf{Iterative Frame Selection.} The frame selection process employs a three-stage pipeline that adaptively gathers relevant information based on the agent's current understanding. Initially, the system uniformly samples $N$ frames $\mathcal{F}_0$ from the video to construct a baseline knowledge graph. From these frames' captions, the LLM generates a preliminary answer $a_0$ and confidence score $c_0$, establishing a foundational understanding of the video content. 

When the confidence score falls below the threshold $\tau = 3$, indicating insufficient information, the system activates additional retrieval stages. The second stage implements a sophisticated retrieval mechanism that leverages both the graph structure and visual features. Frame scoring is computed through a weighted combination:

\begin{equation}
S(f) = \alpha \cdot s_{\text{graph}}(f) + \beta \cdot s_{\text{visual}}(f) + \gamma \cdot s_{\text{temporal}}(f)
\end{equation}

where $s_{\text{graph}}$, $s_{\text{visual}}$, and $s_{\text{temporal}}$ represent normalized scores (range [0,1]) evaluating graph relation relevance, visual similarity, and temporal coherence respectively. The weights are empirically set to $\alpha = 0.5$, $\beta = 0.3$, and $\gamma = 0.2$ to balance different information sources. If confidence remains insufficient after the second stage, a third retrieval iteration is performed with expanded context, enabling exploration of more distant temporal relations. Each retrieval stage maintains efficiency by limiting additional frames to 3, ensuring focused information gathering while avoiding redundant processing.

\noindent\textbf{Question Answering Process.} The question answering process integrates the graph representation with a sophisticated multi-round reasoning framework. In each iteration, the process consists of three key steps: state evaluation, action determination, and state updating. During state evaluation, the LLM analyzes the current state $s_t$ alongside the graph structure $G$, employing chain-of-thought prompting to generate predictions and self-reflection to assess confidence on a three-level scale: insufficient (1), partial (2), and sufficient information (3).

The action determination step follows the formula \begin{equation}
P(a|q,G,F) = \text{LLM}(\text{Prompt}(q, G_e, G_r, G_t))
\end{equation}
where $G_e$, $G_r$, and $G_t$ represent entity, relation, and temporal information respectively. When confidence reaches level 3, the system proceeds to answer generation. Otherwise, it initiates segment-aware information gathering, using the graph structure to identify relevant video segments based on temporal relations and entity states.

The state updating process integrates newly retrieved information into both the current state and graph structure through $G_{t+1} = U(G_t, F_{t+1}, R_{t+1})$. This segment-aware approach is particularly effective for complex temporal reasoning tasks, such as tracking state changes or understanding causal relations. The process continues iteratively until either reaching sufficient confidence or the maximum iteration limit.

This multi-round approach offers significant advantages over single-shot methods by enabling focused information gathering and maintaining computational efficiency through selective processing. Integrating graph structure with iterative reasoning enhances the system's capability in handling complex temporal queries, particularly those requiring an understanding of long-term dependencies and subtle state changes across different time spans.

\begin{table*}[t]
    \begin{minipage}{0.43\textwidth}
        \centering
        \caption{\emph{Results on EgoSchema compared
to public models.} Full-set results are obtained from the official leaderboard.}
        \resizebox{\textwidth}{!}{
            \setlength\tabcolsep{2pt}
            \renewcommand\arraystretch{1.03}
            \begin{tabular}{rl|ccc}
                \toprule
                \multicolumn{2}{c|}{Method} & Frames & Subset & \textbf{Full} \\ 
                \midrule
                FrozenBiLM~\cite{yang2022frozenbilm} \!&\pub{NeurIPS2022} & 90 & - & 26.9 \\
                InternVideo~\cite{wang2022internvideo} \!&\pub{arXiv2022.12}& 90 & - & 32.1 \\
                ImageViT~\cite{papalampidi2023simple} \!&\pub{CVPR2024}& 16 & 40.8 & 30.9 \\
                ShortViViT$_{loc}$~\cite{papalampidi2023simple} \!&\pub{CVPR2024} & 32 & 49.6 & 31.3 \\  
                LongViViT~\cite{papalampidi2023simple} \!&\pub{CVPR2024} & 256 & 56.8 & 33.3 \\  
                SeViLA~\cite{yu2023self} \!&\pub{NeurIPS2023}& 32 & 25.7 & 22.7 \\ 
                Vamos~\cite{Wang2024Vamos} \!&\pub{ECCV2024}  & - & - & 48.3 \\
                LLoVi~\cite{zhang2023simple} \!&\pub{ACL2024}  & 180 & 57.6 & 50.3 \\
                MC-ViT-L~\cite{balavzevic2024memory} \!&\pub{ICML2024}& 128+ & 62.6 & 44.4 \\
                \midrule 
                VideoAgent~\cite{wang2025videoagent}&\footnotesize{(\texttt{base})} & 8.4 & 60.2 & 54.1 \\ 
                GraphVideoAgent&\footnotesize{(\textbf{\texttt{ours}})} & \textbf{8.2} & \textbf{62.7} & \textbf{56.3} \\
                \bottomrule
            \end{tabular}
        }

        \label{tab:ego-public}
    \end{minipage}
    \hfill
    \begin{minipage}{0.47\textwidth}
        \centering
        \caption{\emph{Results on EgoSchema compared to large-scale proprietary models.}}
        \resizebox{\textwidth}{!}{
            \setlength\tabcolsep{4pt}
            \renewcommand\arraystretch{1.03}
            \begin{tabular}{rl|cc}
                \toprule
                \multicolumn{2}{c|}{Model} & Subset & \textbf{Full}\\ 
                \midrule
                Random Chance && 20.0 & 20.0 \\
                Bard only (blind)~\cite{balavzevic2024memory} \!&\pub{2023.3} & 27.0 & 33.2 \\
                Bard + ImageViT~\cite{papalampidi2023simple} \!&\pub{2023.3} & 35.0 & 35.0\\
                Bard + ShortViViT~\cite{papalampidi2023simple} \!&\pub{2023.3} & 42.0 & 36.2\\
                Bard + PALI~\cite{papalampidi2023simple} \!&\pub{2023.3}  & 44.8 & 39.2 \\
                GPT-4 Turbo (blind)~\cite{balavzevic2024memory} \!&\pub{2023.3}& 31.0 & 30.8 \\
                GPT-4V~\cite{balavzevic2024memory} \!&\pub{2023.3} & 63.5 & 55.6 \\
                Gemini 1.0 Pro~\cite{team2023gemini} \!&\pub{2023.12}& - & 55.7 \\
                \midrule
                VideoAgent~\cite{wang2025videoagent}&\footnotesize{(\texttt{base})} & 60.2 & 54.1 \\ 
                GraphVideoAgent&\footnotesize{(\textbf{\texttt{ours}})} & \textbf{62.7} & \textbf{56.3} \\
                \bottomrule
            \end{tabular}
        }
        \label{tab:ego-private}
    \end{minipage}
\end{table*}

\begin{table*}[t]
    \centering
    \footnotesize
                \caption{\emph{Results on NExT-QA compared to the state of the art.} C, T, and D are causal, temporal, and descriptive subsets, respectively.}
    \resizebox{0.8\textwidth}{!}{
        \setlength\tabcolsep{2pt}
        \renewcommand\arraystretch{0.9}
        \begin{tabular}{rl|cccc|ccc}
            \toprule
            & & \multicolumn{4}{c|}{Val} & \multicolumn{3}{c}{ATP-hard subset}  \\
            \multicolumn{2}{c|}{\multirow{-2}{*}{Methods}}  & Acc@C & Acc@T & Acc@D &  Acc@All & Acc@C & Acc@T &  Acc@All  \\ 
            \midrule
            \multicolumn{9}{c}{\textit{Supervised}} \\
            VFC~\cite{yang2021just}\!&\pub{ICCV2021} & 49.6 & 51.5 & 63.2 & 52.3 & - & - & - \\
            ATP~\cite{buch2022revisiting}\!&\pub{CVPR2022} & 53.1 & 50.2 & 66.8 & 54.3 & 38.4 & 36.5 & 38.8 \\
            MIST~\cite{gao2023mist} \!&\pub{CVPR2023} & 54.6 &  56.6 & 66.9 &  57.2 & - & - & -\\
            GF~\cite{bai2024glance} \!&\pub{NeurIPS2023}  & 56.9 & 57.1 &  70.5 & 58.8 &48.7 & 50.3 & 49.3 \\
            CoVGT~\cite{xiaovgt} \!&\pub{TPAMI2023}  &59.7 & 58.0& 69.9 & 60.7 & - &-&- \\
            SeViT~\cite{kim2023semi} \!&\pub{arXiv2023.1} & 54.0 & 54.1 & 71.3 & 56.7 & 43.3 & 46.5 & - \\
            HiTeA~\cite{ye2023hitea} \!&\pub{ICCV2023} & 62.4 & 58.3 & 75.6 & 63.1 & 47.8 & 48.6 &  - \\
            \midrule
            \multicolumn{9}{c}{\textit{Zero-shot}} \\
            VFC~\cite{momeni2023verbs} \!&\pub{ICCV2023} & 51.6 & 45.4 & 64.1 & 51.5 & 32.2 & 30.0 & 31.4 \\
            InternVideo~\cite{wang2022internvideo} \!&\pub{arXiv2022.12}& 43.4 & 48.0 & 65.1 & 49.1 & - & - & - \\
            AssistGPT~\cite{gao2023assistgpt} \!&\pub{arXiv2023.6} & 60.0 & 51.4 & 67.3 & 58.4 & - & - & - \\
            ViperGPT~\cite{suris2023vipergpt} \!&\pub{ICCV2023}& - & - & - & 60.0 & - & - & - \\
            SeViLA~\cite{yu2023self} \!&\pub{NeurIPS2023}& 61.3 & 61.5 & 75.6 & 63.6 & - & - & -  \\
            LLoVi~\cite{zhang2023simple} \!&\pub{arXiv2024.2}  & 69.5 & 61.0 & 75.6 & 67.7 & - & - & - \\
            \midrule
            VideoAgent~\cite{wang2025videoagent}&\footnotesize{(\texttt{base})} & 72.7 & 64.5 & 81.1 & 71.3 & 57.8 & 58.8 & 58.4  \\
            GraphVideoAgent&\footnotesize{(\textbf{\texttt{ours}})} & \textbf{74.6} & \textbf{65.2} & \textbf{83.5} & \textbf{73.3} & \textbf{59.2} & \textbf{60.1} & \textbf{59.7}  \\
            \bottomrule
        \end{tabular}
    }
    \label{tab:nextqa}
\end{table*}

\section{Experiments}

\subsection{Experimental Setup}
We conduct extensive evaluations of GraphVideoAgent on two challenging video understanding benchmarks that test different aspects of video comprehension capabilities: EgoSchema~\cite{mangalam2023egoschema} and NExT-QA~\cite{xiao2021next}. EgoSchema presents a particularly challenging test bed, containing multiple-choice questions based on egocentric videos that require understanding of first-person perspectives and complex human-object interactions. The benchmark provides both a full test set and a public subset, enabling comprehensive evaluation across different data regimes. NExT-QA complements this with a diverse set of complex temporal, causal, and descriptive questions that demand sophisticated reasoning about video content over time. Our implementation leverages several powerful foundation models: EVA-CLIP-8B-plus \cite{sun2023eva} operating at 448×448 resolution for high-quality frame feature extraction, LaViLa \cite{lavila} specifically for egocentric video captioning to handle the unique challenges of first-person viewpoints, and GPT-4 serving as the primary LLM agent for reasoning and answer generation. To establish the effectiveness of our approach, we conduct comprehensive comparisons against an extensive set of state-of-the-art models, including advanced video understanding systems like LLoVi \cite{llovi} and MC-ViT-L \cite{balavzevic2024memory}, specialized models like SeViLA \cite{sevila}, and cutting-edge proprietary models such as GPT-4V \cite{openai2023gpt4} and Gemini 1.0 Pro \cite{team2023gemini}. We also collect a multi-entity hour-long video dataset to test its multi-entity understanding performance.

\subsection{Main Results}
\noindent\textbf{EgoSchema Results.} Our experimental results, as detailed in Table \ref{tab:ego-public} and \ref{tab:ego-private}, demonstrate GraphVideoAgent's exceptional performance across multiple evaluation settings. The model achieves state-of-the-art results on both the full test set with 56.3\% accuracy and the public subset with 62.7\% accuracy, marking significant improvements over our already strong baseline VideoAgent (54.1\% and 60.2\% respectively). Particularly noteworthy is our model's ability to outperform sophisticated proprietary systems including GPT-4V (55.6\%) and Gemini 1.0 Pro (55.7\%) on the full test set. What makes these results especially impressive is the remarkable efficiency of our approach - GraphVideoAgent requires only 8.2 frames per video for analysis, while competing methods demand substantially more computational resources, processing between 32 to 256 frames (for instance, LongViViT \cite{papalampidi2024simple} processes 256 frames and LLoVi \cite{llovi} requires 180 frames). The consistent performance improvement over VideoAgent observed across both the subset and full test scenarios (62.7\% vs 56.3\%) provides strong empirical validation for the effectiveness of our graph-based entity tracking approach. The performance difference between the subset (60.2\%) and full test set (54.1\%) reflects the distinct characteristics of our evaluation protocol. The subset, comprising approximately 10\% of the full dataset, represents a carefully curated selection of videos that enables detailed analysis of specific video understanding capabilities. This controlled subset allows for more thorough examination of the model's performance on specific video understanding challenges, while the full test set provides a comprehensive evaluation across a broader range of scenarios.

\input{tab/ablation}

\noindent\textbf{NExT-QA Results.} The results presented in Table \ref{tab:nextqa} reveal Graph-VideoAgent's comprehensive superiority across all question categories in the NExT-QA benchmark. With an average of merely 8.1 frames used per video, our model achieves an outstanding 73.3\% overall accuracy, substantially surpassing both traditional supervised methods (with the previous best being HiTeA at 63.1\%) and zero-shot approaches (where LLoVi held the previous record at 67.7\%). The performance improvements are particularly pronounced across different question types: causal questions (74.6\%), temporal questions (65.2\%), and descriptive questions (83.5\%), demonstrating consistent and significant gains over the VideoAgent baseline across all reasoning categories. Our model's robustness is further validated by its performance on the ATP-hard subset, a particularly challenging collection of questions requiring complex reasoning, where we achieve 59.7\% accuracy compared to VideoAgent's 58.4\%. These results provide compelling evidence for the effectiveness of our graph-based approach in handling sophisticated reasoning tasks across diverse question types.

\subsection{Ablation Studyies of Graph VideoAgent}

To systematically evaluate the efficacy of our proposed approach, we conduct extensive ablation experiments across three critical dimensions: language model selection, graph component contribution, and entity scaling capabilities.

\noindent\textbf{LLM Ablation.} Our investigation into the impact of language model selection reveals significant performance variations. As illustrated in Table \ref{tab:llm}, GPT-4 demonstrates superior performance with an accuracy of 62.7\%, substantially exceeding GPT-3.5 (49.8\%) and other contemporary language models. Within the 70B parameter regime, Llama3-70B achieves 50.1\% accuracy, markedly outperforming Mistral-8x7B (39.6\%). This pronounced disparity between architectures of comparable scale underscores the critical importance of model architecture and pre-training methodology beyond mere parameter count. The substantial performance delta between GPT-3.5 and GPT-4 (12.9 percentage points) suggests that advanced language models possess enhanced capabilities in comprehending complex video-language relationships and executing sophisticated multi-entity reasoning tasks.

\noindent\textbf{Graph Component Ablation.} Our architectural analysis examines the relative contributions of three fundamental components: entity relations, temporal tracking, and multi-dimension structure. The empirical results presented in Table \ref{tab:graph_ablation} demonstrate that the elimination of any component results in consistent performance degradation across all evaluation metrics. The complete model architecture achieves optimal performance on both EgoSchema (56.3\%) and NExT-QA (73.3\%). Notably, the ablation of multi-dimension structure induces the most significant performance deterioration (52.1\% on EgoSchema), emphasizing its fundamental role in video understanding. The comparatively moderate impact observed from removing entity relations (53.8\%) and temporal tracking (54.2\%) indicates that while these components enhance overall system performance, the multi-dimension structure constitutes the cornerstone of our methodology's effectiveness. This observation aligns with our theoretical framework, suggesting that the graph's capacity to encode global contextual information and relational dependencies is essential for sophisticated video comprehension tasks.

\noindent\textbf{Entity Scale Analysis.} Our investigation into scalability characteristics reveals compelling insights into the model's performance across varying entity complexities. The evaluation framework utilizes 50 multiple-choice questions, each containing four options with one correct answer. The experimental results presented in Table \ref{tab: multi-entity-mcq} demonstrate that while performance exhibits an expected decline with increasing entity count (from 64.0\% with 2-3 entities to 52.0\% with 7+ entities), our methodology consistently surpasses the VideoAgent baseline across all scale regimes. The performance differential amplifies with increasing entity complexity, expanding from 4 percentage points in scenarios with 2-3 entities to 8 percentage points in cases involving 7+ entities, indicating superior scalability of our graph-based architecture. This widening performance gap provides empirical evidence that our graph structure more effectively manages the increased computational and representational demands of complex multi-entity scenarios. The observed gradual performance degradation
(64.0\% → 58.0\% → 52.0\%) exhibits notably more graceful
scaling characteristics compared to the baseline, suggesting
enhanced potential for adaptation to increasingly complex
visual scenarios.

\begin{figure*}[t]
\centering
\includegraphics[width=\linewidth]{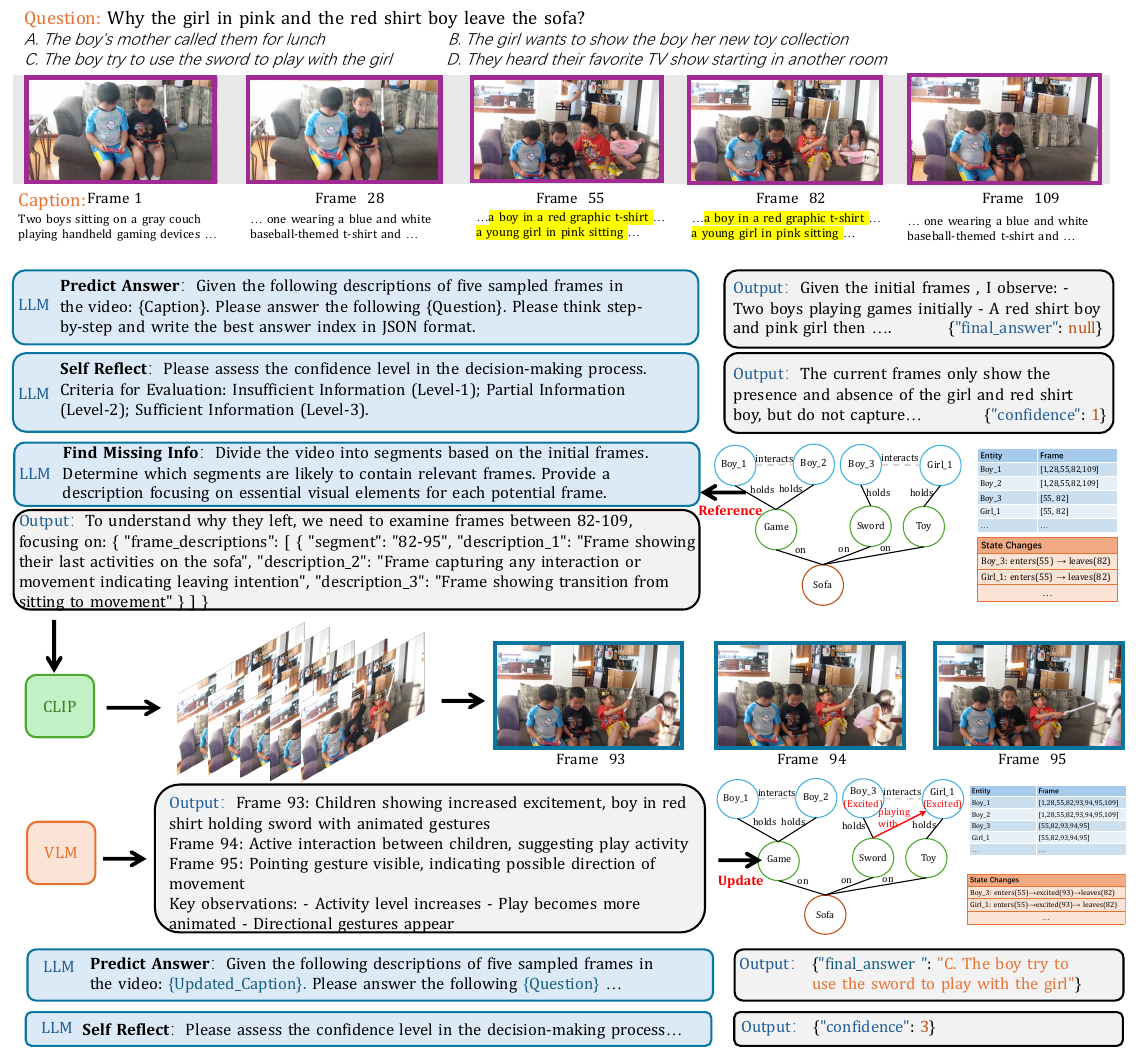}
\caption{The figure demonstrates a complete analysis pipeline for understanding why children leave a sofa. Starting with initial frames and entity recognition (top), the system employs three key components: (1) The LLM agent for iterative reasoning through "Predict Answer", "Self Reflect", and "Find Missing Info" stages, (2) Foundation tools including CLIP for frame retrieval and VLM for frame captioning (middle), and (3) A dynamic graph structure (right) that tracks entities (Boys 1-3, Girl 1), their relations (interacts, holds), objects (Game, Sword, Toy), and state changes over time. The graph is iteratively updated as new information is discovered, enabling accurate tracking of interactions ("excited", "playing") and state transitions that lead to the final answer.}
\label{fig:case1}
\end{figure*}

\begin{figure*}[t]
\centering
\includegraphics[width=0.8\linewidth]{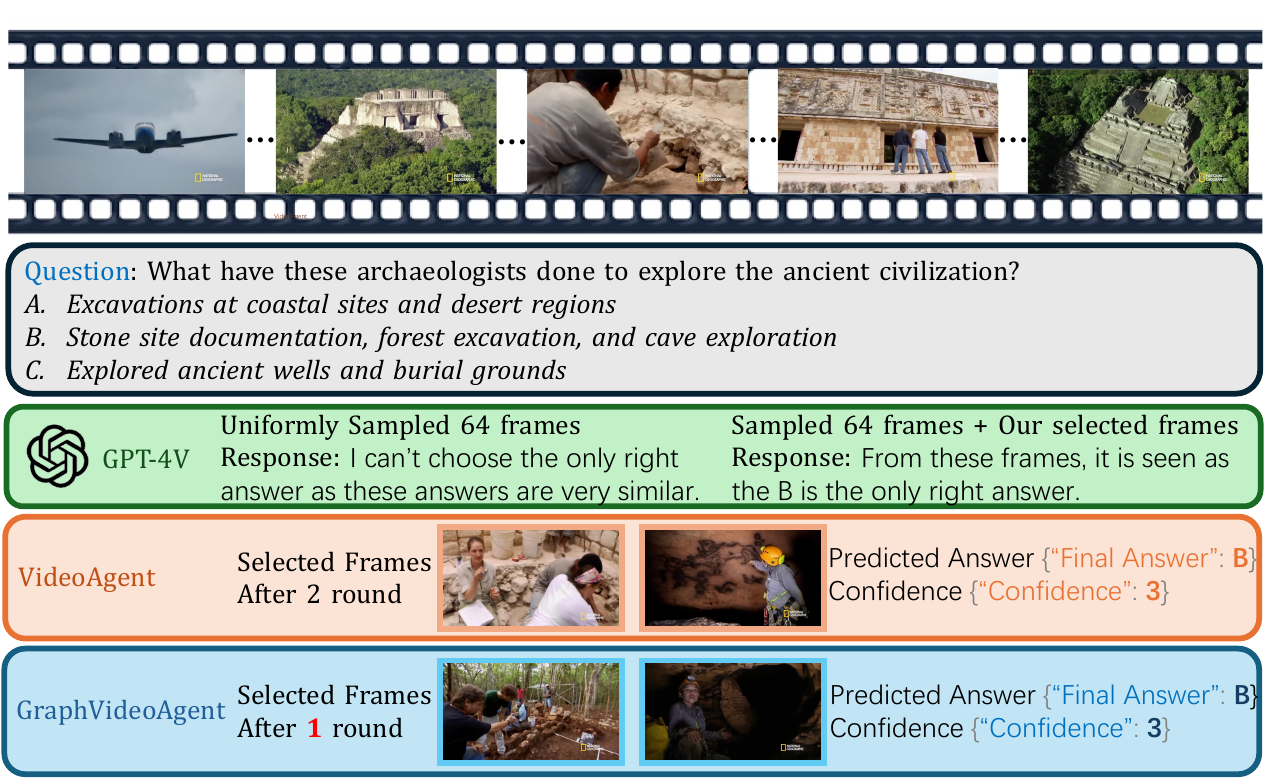}
\caption{The example demonstrates GraphVideoAgent's improved frame selection efficiency compared to VideoAgent and GPT-4V in understanding archaeological exploration activities. When analyzing a question about archaeologists' methods of exploring ancient civilizations, GPT-4V with uniform sampling of 64 frames fails to make a definitive choice, indicating insufficient understanding. However, with targeted frame selection, both VideoAgent and GraphVideoAgent successfully identify option B as the correct answer, with GraphVideoAgent requiring only one round of selection compared to VideoAgent's two rounds. This highlights how GraphVideoAgent's graph-based approach enables more efficient information gathering while maintaining high confidence in predictions. 
}
\label{fig:case2}
\end{figure*}

\subsection{Case Studies}
We present several case studies to demonstrate the capability of GraphVideoAgent in understanding long-form videos.

\noindent\textbf{Complex Interaction Analysis.} In Figure \ref{fig:case1}, we demonstrate GraphVideoAgent's sophisticated temporal and behavioral reasoning capabilities through a scene depicting children leaving a sofa. While initial frame analysis provides basic information about entity presence, our system employs a three-stage confidence assessment framework that reveals critical interaction patterns. Upon detecting information gaps in the initial frames, GraphVideoAgent leverages its graph-structured architecture to identify and analyze pivotal behavioral changes. The system successfully captures the emergence of dynamic interactions, particularly the boy's engagement with the sword, and constructs a detailed temporal map of state transitions. This analysis not only identifies what occurred but also establishes clear causal relations between the introduction of the sword play and subsequent changes in children's behavior, demonstrating the system's advanced understanding of social interaction dynamics. The system's entity tracking mechanism maintains precise temporal records (e.g., "Boy 1 [1,28,55,82,109]") and state changes, enabling reconstruction of complex behavioral sequences. This granular tracking, combined with the system's ability to recognize emotional states and their transitions, allows for nuanced interpretation of social cues and behavioral motivations that traditional frame-based analysis might miss.

\noindent\textbf{Multi-Entity Scene Understanding.} Figure \ref{fig:case2} showcases 
 Graph-VideoAgent's enhanced processing capabilities in complex scenarios involving multiple entities and evolving relations. Through its dual-layer graph representation, the system maintains comprehensive tracking of entity states, spatial relations, and temporal transitions. The initial reference graph establishes foundational relations among three boys, a girl, and various objects, while the dynamic graph layer captures evolving emotional states and behavioral changes. This sophisticated approach enables single-round scene comprehension, contrasting favorably with VideoAgent's two-round requirement for similar understanding. The system's ability to identify the sword play as a pivotal interaction demonstrates its effectiveness in capturing cause-and-effect relations. Furthermore, the graph-structured methodology shows significant efficiency advantages, requiring fewer processing iterations while maintaining high accuracy in understanding complex social dynamics. Notably, the system's hierarchical processing approach demonstrates superior performance compared to traditional methods like GPT-4V's uniform sampling of 64 frames, which fails to reach definitive conclusions. The graph structure's ability to track both explicit physical relations (e.g., "holds", "on", "interacts") and implicit social dynamics enables comprehensive scene understanding while maintaining computational efficiency. This multi-layered approach to scene analysis represents a significant advance in automated understanding of complex social interactions in video content.

\section{Conclusion}

This work introduces GraphVideoAgent, a novel approach to long-form video understanding that leverages dynamic entity relation graphs to enhance temporal reasoning. Our method demonstrates that explicitly modelling entity relations across frames through a graph structure can substantially improve video understanding while maintaining computational efficiency. By achieving state-of-the-art performance on both EgoSchema (56.3\%) and NExT-QA (73.3\%) with remarkably few frames (8.2 and 8.1 on average), GraphVideoAgent validates the effectiveness of structured semantic memory in video understanding tasks. Looking ahead, several promising directions emerge for extending this work. The graph structure could be enhanced to capture more complex entity relations and multi-modal information, while more efficient graph construction mechanisms could enable real-time applications. Moreover, our work highlights the value of incorporating structured representations to make video understanding systems more efficient. Future work could enhance the graph structure to handle more complex entity relations and multi-modal information while developing more efficient graph construction mechanisms that enable real-time applications.

\bibliographystyle{ACM-Reference-Format}
\bibliography{main}


\begin{thebibliography}{66}


\ifx \showCODEN    \undefined \def \showCODEN     #1{\unskip}     \fi
\ifx \showDOI      \undefined \def \showDOI       #1{#1}\fi
\ifx \showISBNx    \undefined \def \showISBNx     #1{\unskip}     \fi
\ifx \showISBNxiii \undefined \def \showISBNxiii  #1{\unskip}     \fi
\ifx \showISSN     \undefined \def \showISSN      #1{\unskip}     \fi
\ifx \showLCCN     \undefined \def \showLCCN      #1{\unskip}     \fi
\ifx \shownote     \undefined \def \shownote      #1{#1}          \fi
\ifx \showarticletitle \undefined \def \showarticletitle #1{#1}   \fi
\ifx \showURL      \undefined \def \showURL       {\relax}        \fi
\providecommand\bibfield[2]{#2}
\providecommand\bibinfo[2]{#2}
\providecommand\natexlab[1]{#1}
\providecommand\showeprint[2][]{arXiv:#2}

\bibitem[Anil et~al\mbox{.}(2023)]%
        {team2023gemini}
\bibfield{author}{\bibinfo{person}{Rohan Anil}, \bibinfo{person}{Sebastian Borgeaud}, \bibinfo{person}{Yonghui Wu}, \bibinfo{person}{Jean{-}Baptiste Alayrac}, \bibinfo{person}{Jiahui Yu}, \bibinfo{person}{Radu Soricut}, \bibinfo{person}{Johan Schalkwyk}, \bibinfo{person}{Andrew~M. Dai}, \bibinfo{person}{Anja Hauth}, \bibinfo{person}{Katie Millican}, \bibinfo{person}{David Silver}, \bibinfo{person}{Slav Petrov}, \bibinfo{person}{Melvin Johnson}, \bibinfo{person}{Ioannis Antonoglou}, \bibinfo{person}{Julian Schrittwieser}, \bibinfo{person}{Amelia Glaese}, \bibinfo{person}{Jilin Chen}, \bibinfo{person}{Emily Pitler}, \bibinfo{person}{Timothy~P. Lillicrap}, \bibinfo{person}{Angeliki Lazaridou}, \bibinfo{person}{Orhan Firat}, \bibinfo{person}{James Molloy}, \bibinfo{person}{Michael Isard}, \bibinfo{person}{Paul~Ronald Barham}, \bibinfo{person}{Tom Hennigan}, \bibinfo{person}{Benjamin Lee}, \bibinfo{person}{Fabio Viola}, \bibinfo{person}{Malcolm Reynolds}, \bibinfo{person}{Yuanzhong Xu}, \bibinfo{person}{Ryan
  Doherty}, \bibinfo{person}{Eli Collins}, \bibinfo{person}{Clemens Meyer}, \bibinfo{person}{Eliza Rutherford}, \bibinfo{person}{Erica Moreira}, \bibinfo{person}{Kareem Ayoub}, \bibinfo{person}{Megha Goel}, \bibinfo{person}{George Tucker}, \bibinfo{person}{Enrique Piqueras}, \bibinfo{person}{Maxim Krikun}, \bibinfo{person}{Iain Barr}, \bibinfo{person}{Nikolay Savinov}, \bibinfo{person}{Ivo Danihelka}, \bibinfo{person}{Becca Roelofs}, \bibinfo{person}{Ana{\"{\i}}s White}, \bibinfo{person}{Anders Andreassen}, \bibinfo{person}{Tamara von Glehn}, \bibinfo{person}{Lakshman Yagati}, \bibinfo{person}{Mehran Kazemi}, \bibinfo{person}{Lucas Gonzalez}, \bibinfo{person}{Misha Khalman}, \bibinfo{person}{Jakub Sygnowski}, {and} \bibinfo{person}{et al.}} \bibinfo{year}{2023}\natexlab{}.
\newblock \showarticletitle{Gemini: {A} Family of Highly Capable Multimodal Models}.
\newblock \bibinfo{journal}{\emph{CoRR}}  \bibinfo{volume}{abs/2312.11805} (\bibinfo{year}{2023}).
\newblock
\urldef\tempurl%
\url{https://doi.org/10.48550/ARXIV.2312.11805}
\showDOI{\tempurl}
\showeprint[arXiv]{2312.11805}


\bibitem[Baddeley et~al\mbox{.}(2020)]%
        {baddeley2020memory}
\bibfield{author}{\bibinfo{person}{Alan Baddeley}, \bibinfo{person}{Michael~W. Eysenck}, {and} \bibinfo{person}{Michael~C. Anderson}.} \bibinfo{year}{2020}\natexlab{}.
\newblock \bibinfo{booktitle}{\emph{Memory} (\bibinfo{edition}{3} ed.)}.
\newblock \bibinfo{publisher}{Routledge}.
\newblock
\urldef\tempurl%
\url{https://doi.org/10.4324/9780429449642}
\showDOI{\tempurl}


\bibitem[Bai et~al\mbox{.}(2024)]%
        {bai2024glance}
\bibfield{author}{\bibinfo{person}{Ziyi Bai}, \bibinfo{person}{Ruiping Wang}, {and} \bibinfo{person}{Xilin Chen}.} \bibinfo{year}{2024}\natexlab{}.
\newblock \showarticletitle{Glance and Focus: Memory Prompting for Multi-Event Video Question Answering}.
\newblock \bibinfo{journal}{\emph{Advances in Neural Information Processing Systems}}  \bibinfo{volume}{36} (\bibinfo{year}{2024}).
\newblock


\bibitem[Bala{\v{z}}evi{\'c} et~al\mbox{.}(2024)]%
        {balavzevic2024memory}
\bibfield{author}{\bibinfo{person}{Ivana Bala{\v{z}}evi{\'c}}, \bibinfo{person}{Yuge Shi}, \bibinfo{person}{Pinelopi Papalampidi}, \bibinfo{person}{Rahma Chaabouni}, \bibinfo{person}{Skanda Koppula}, {and} \bibinfo{person}{Olivier~J H{\'e}naff}.} \bibinfo{year}{2024}\natexlab{}.
\newblock \showarticletitle{Memory Consolidation Enables Long-Context Video Understanding}.
\newblock \bibinfo{journal}{\emph{arXiv preprint arXiv:2402.05861}} (\bibinfo{year}{2024}).
\newblock


\bibitem[Brown et~al\mbox{.}(2020)]%
        {brown2020language}
\bibfield{author}{\bibinfo{person}{Tom Brown}, \bibinfo{person}{Benjamin Mann}, \bibinfo{person}{Nick Ryder}, \bibinfo{person}{Melanie Subbiah}, \bibinfo{person}{Jared~D Kaplan}, \bibinfo{person}{Prafulla Dhariwal}, \bibinfo{person}{Arvind Neelakantan}, \bibinfo{person}{Pranav Shyam}, \bibinfo{person}{Girish Sastry}, \bibinfo{person}{Amanda Askell}, {et~al\mbox{.}}} \bibinfo{year}{2020}\natexlab{}.
\newblock \showarticletitle{Language models are few-shot learners}.
\newblock \bibinfo{journal}{\emph{Advances in neural information processing systems}}  \bibinfo{volume}{33} (\bibinfo{year}{2020}), \bibinfo{pages}{1877--1901}.
\newblock


\bibitem[Buch et~al\mbox{.}(2022)]%
        {buch2022revisiting}
\bibfield{author}{\bibinfo{person}{Shyamal Buch}, \bibinfo{person}{Crist{\'o}bal Eyzaguirre}, \bibinfo{person}{Adrien Gaidon}, \bibinfo{person}{Jiajun Wu}, \bibinfo{person}{Li Fei-Fei}, {and} \bibinfo{person}{Juan~Carlos Niebles}.} \bibinfo{year}{2022}\natexlab{}.
\newblock \showarticletitle{Revisiting the" video" in video-language understanding}. In \bibinfo{booktitle}{\emph{Proceedings of the IEEE/CVF conference on computer vision and pattern recognition}}. \bibinfo{pages}{2917--2927}.
\newblock


\bibitem[Chu et~al\mbox{.}(2024)]%
        {chu20243d}
\bibfield{author}{\bibinfo{person}{Meng Chu}, \bibinfo{person}{Xuan Zhang}, \bibinfo{person}{Zhedong Zheng}, {and} \bibinfo{person}{Tat-Seng Chua}.} \bibinfo{year}{2024}\natexlab{}.
\newblock \showarticletitle{3D-TAFS: A Training-free Framework for 3D Affordance Segmentation}.
\newblock \bibinfo{journal}{\emph{arXiv preprint arXiv:2409.10078}} (\bibinfo{year}{2024}).
\newblock


\bibitem[Chu et~al\mbox{.}(2025)]%
        {chu2025towards}
\bibfield{author}{\bibinfo{person}{Meng Chu}, \bibinfo{person}{Zhedong Zheng}, \bibinfo{person}{Wei Ji}, \bibinfo{person}{Tingyu Wang}, {and} \bibinfo{person}{Tat-Seng Chua}.} \bibinfo{year}{2025}\natexlab{}.
\newblock \showarticletitle{Towards natural language-guided drones: GeoText-1652 benchmark with spatial relation matching}. In \bibinfo{booktitle}{\emph{European Conference on Computer Vision}}. Springer, \bibinfo{pages}{213--231}.
\newblock


\bibitem[Cohendet et~al\mbox{.}(2019)]%
        {cohendet2019videomem}
\bibfield{author}{\bibinfo{person}{Romain Cohendet}, \bibinfo{person}{Claire-H{\'e}l{\`e}ne Demarty}, \bibinfo{person}{Ngoc~QK Duong}, {and} \bibinfo{person}{Martin Engilberge}.} \bibinfo{year}{2019}\natexlab{}.
\newblock \showarticletitle{VideoMem: Constructing, analyzing, predicting short-term and long-term video memorability}. In \bibinfo{booktitle}{\emph{Proceedings of the IEEE/CVF International Conference on Computer Vision}}. \bibinfo{pages}{2531--2540}.
\newblock


\bibitem[Dubey et~al\mbox{.}(2024)]%
        {dubey2024llama}
\bibfield{author}{\bibinfo{person}{Abhimanyu Dubey}, \bibinfo{person}{Abhinav Jauhri}, \bibinfo{person}{Abhinav Pandey}, \bibinfo{person}{Abhishek Kadian}, \bibinfo{person}{Ahmad Al-Dahle}, \bibinfo{person}{Aiesha Letman}, \bibinfo{person}{Akhil Mathur}, \bibinfo{person}{Alan Schelten}, \bibinfo{person}{Amy Yang}, \bibinfo{person}{Angela Fan}, {et~al\mbox{.}}} \bibinfo{year}{2024}\natexlab{}.
\newblock \showarticletitle{The llama 3 herd of models}.
\newblock \bibinfo{journal}{\emph{arXiv preprint arXiv:2407.21783}} (\bibinfo{year}{2024}).
\newblock


\bibitem[Gao et~al\mbox{.}(2023a)]%
        {gao2023assistgpt}
\bibfield{author}{\bibinfo{person}{Difei Gao}, \bibinfo{person}{Lei Ji}, \bibinfo{person}{Luowei Zhou}, \bibinfo{person}{Kevin~Qinghong Lin}, \bibinfo{person}{Joya Chen}, \bibinfo{person}{Zihan Fan}, {and} \bibinfo{person}{Mike~Zheng Shou}.} \bibinfo{year}{2023}\natexlab{a}.
\newblock \showarticletitle{AssistGPT: A General Multi-modal Assistant that can Plan, Execute, Inspect, and Learn}.
\newblock \bibinfo{journal}{\emph{arXiv preprint arXiv:2306.08640}} (\bibinfo{year}{2023}).
\newblock


\bibitem[Gao et~al\mbox{.}(2023b)]%
        {gao2023mist}
\bibfield{author}{\bibinfo{person}{Difei Gao}, \bibinfo{person}{Luowei Zhou}, \bibinfo{person}{Lei Ji}, \bibinfo{person}{Linchao Zhu}, \bibinfo{person}{Yi Yang}, {and} \bibinfo{person}{Mike~Zheng Shou}.} \bibinfo{year}{2023}\natexlab{b}.
\newblock \showarticletitle{MIST: Multi-modal Iterative Spatial-Temporal Transformer for Long-form Video Question Answering}. In \bibinfo{booktitle}{\emph{Proceedings of the IEEE/CVF Conference on Computer Vision and Pattern Recognition}}. \bibinfo{pages}{14773--14783}.
\newblock


\bibitem[Gao et~al\mbox{.}(2024)]%
        {gao2024graphdreamer}
\bibfield{author}{\bibinfo{person}{Gege Gao}, \bibinfo{person}{Weiyang Liu}, \bibinfo{person}{Anpei Chen}, \bibinfo{person}{Andreas Geiger}, {and} \bibinfo{person}{Bernhard Sch{\"o}lkopf}.} \bibinfo{year}{2024}\natexlab{}.
\newblock \showarticletitle{Graphdreamer: Compositional 3d scene synthesis from scene graphs}. In \bibinfo{booktitle}{\emph{Proceedings of the IEEE/CVF Conference on Computer Vision and Pattern Recognition}}. \bibinfo{pages}{21295--21304}.
\newblock


\bibitem[Ghodrati et~al\mbox{.}(2021)]%
        {ghodrati2021frameexit}
\bibfield{author}{\bibinfo{person}{Amir Ghodrati}, \bibinfo{person}{Babak~Ehteshami Bejnordi}, {and} \bibinfo{person}{Amirhossein Habibian}.} \bibinfo{year}{2021}\natexlab{}.
\newblock \showarticletitle{Frameexit: Conditional early exiting for efficient video recognition}. In \bibinfo{booktitle}{\emph{Proceedings of the IEEE/CVF Conference on Computer Vision and Pattern Recognition}}. \bibinfo{pages}{15608--15618}.
\newblock


\bibitem[Hong et~al\mbox{.}(2024)]%
        {hong2024cogagent}
\bibfield{author}{\bibinfo{person}{Wenyi Hong}, \bibinfo{person}{Weihan Wang}, \bibinfo{person}{Qingsong Lv}, \bibinfo{person}{Jiazheng Xu}, \bibinfo{person}{Wenmeng Yu}, \bibinfo{person}{Junhui Ji}, \bibinfo{person}{Yan Wang}, \bibinfo{person}{Zihan Wang}, \bibinfo{person}{Yuxiao Dong}, \bibinfo{person}{Ming Ding}, {et~al\mbox{.}}} \bibinfo{year}{2024}\natexlab{}.
\newblock \showarticletitle{Cogagent: A visual language model for gui agents}. In \bibinfo{booktitle}{\emph{Proceedings of the IEEE/CVF Conference on Computer Vision and Pattern Recognition}}. \bibinfo{pages}{14281--14290}.
\newblock


\bibitem[Honnibal et~al\mbox{.}(2020)]%
        {spacy2020}
\bibfield{author}{\bibinfo{person}{Matthew Honnibal}, \bibinfo{person}{Ines Montani}, \bibinfo{person}{Sofie Van~Landeghem}, {and} \bibinfo{person}{Adriane Boyd}.} \bibinfo{year}{2020}\natexlab{}.
\newblock \bibinfo{booktitle}{\emph{spaCy: Industrial-strength Natural Language Processing in Python}}.
\newblock
\urldef\tempurl%
\url{https://spacy.io}
\showURL{%
\tempurl}


\bibitem[Hussein et~al\mbox{.}(2019)]%
        {hussein2019videograph}
\bibfield{author}{\bibinfo{person}{Noureldien Hussein}, \bibinfo{person}{Efstratios Gavves}, {and} \bibinfo{person}{Arnold W.~M. Smeulders}.} \bibinfo{year}{2019}\natexlab{}.
\newblock \showarticletitle{VideoGraph: Recognizing Minutes-Long Human Activities in Videos}.
\newblock \bibinfo{journal}{\emph{CoRR}}  \bibinfo{volume}{abs/1905.05143} (\bibinfo{year}{2019}).
\newblock
\showeprint[arXiv]{1905.05143}
\urldef\tempurl%
\url{http://arxiv.org/abs/1905.05143}
\showURL{%
\tempurl}


\bibitem[Islam and Bertasius(2022)]%
        {islam2022long}
\bibfield{author}{\bibinfo{person}{Md~Mohaiminul Islam} {and} \bibinfo{person}{Gedas Bertasius}.} \bibinfo{year}{2022}\natexlab{}.
\newblock \showarticletitle{Long Movie Clip Classification with State-Space Video Models}. In \bibinfo{booktitle}{\emph{Computer Vision - {ECCV} 2022 - 17th European Conference, Tel Aviv, Israel, October 23-27, 2022, Proceedings, Part {XXXV}}} \emph{(\bibinfo{series}{Lecture Notes in Computer Science}, Vol.~\bibinfo{volume}{13695})}, \bibfield{editor}{\bibinfo{person}{Shai Avidan}, \bibinfo{person}{Gabriel~J. Brostow}, \bibinfo{person}{Moustapha Ciss{\'{e}}}, \bibinfo{person}{Giovanni~Maria Farinella}, {and} \bibinfo{person}{Tal Hassner}} (Eds.). \bibinfo{publisher}{Springer}, \bibinfo{pages}{87--104}.
\newblock
\urldef\tempurl%
\url{https://doi.org/10.1007/978-3-031-19833-5\_6}
\showDOI{\tempurl}


\bibitem[Jiang et~al\mbox{.}(2024)]%
        {jiang2024mixtral}
\bibfield{author}{\bibinfo{person}{Albert~Q. Jiang}, \bibinfo{person}{Alexandre Sablayrolles}, \bibinfo{person}{Antoine Roux}, \bibinfo{person}{Arthur Mensch}, \bibinfo{person}{Blanche Savary}, \bibinfo{person}{Chris Bamford}, \bibinfo{person}{Devendra~Singh Chaplot}, \bibinfo{person}{Diego de Las~Casas}, \bibinfo{person}{Emma~Bou Hanna}, \bibinfo{person}{Florian Bressand}, \bibinfo{person}{Gianna Lengyel}, \bibinfo{person}{Guillaume Bour}, \bibinfo{person}{Guillaume Lample}, \bibinfo{person}{L{\'{e}}lio~Renard Lavaud}, \bibinfo{person}{Lucile Saulnier}, \bibinfo{person}{Marie{-}Anne Lachaux}, \bibinfo{person}{Pierre Stock}, \bibinfo{person}{Sandeep Subramanian}, \bibinfo{person}{Sophia Yang}, \bibinfo{person}{Szymon Antoniak}, \bibinfo{person}{Teven~Le Scao}, \bibinfo{person}{Th{\'{e}}ophile Gervet}, \bibinfo{person}{Thibaut Lavril}, \bibinfo{person}{Thomas Wang}, \bibinfo{person}{Timoth{\'{e}}e Lacroix}, {and} \bibinfo{person}{William~El Sayed}.} \bibinfo{year}{2024}\natexlab{}.
\newblock \showarticletitle{Mixtral of Experts}.
\newblock \bibinfo{journal}{\emph{CoRR}}  \bibinfo{volume}{abs/2401.04088} (\bibinfo{year}{2024}).
\newblock
\urldef\tempurl%
\url{https://doi.org/10.48550/ARXIV.2401.04088}
\showDOI{\tempurl}
\showeprint[arXiv]{2401.04088}


\bibitem[Jin et~al\mbox{.}(2023)]%
        {jin2023chatunivi}
\bibfield{author}{\bibinfo{person}{Peng Jin}, \bibinfo{person}{Ryuichi Takanobu}, \bibinfo{person}{Caiwan Zhang}, \bibinfo{person}{Xiaochun Cao}, {and} \bibinfo{person}{Li Yuan}.} \bibinfo{year}{2023}\natexlab{}.
\newblock \bibinfo{title}{Chat-UniVi: Unified Visual Representation Empowers Large Language Models with Image and Video Understanding}.
\newblock
\newblock
\showeprint[arxiv]{2311.08046}~[cs.CV]


\bibitem[Jing et~al\mbox{.}(2020)]%
        {jing2020visual}
\bibfield{author}{\bibinfo{person}{Chenchen Jing}, \bibinfo{person}{Yuwei Wu}, \bibinfo{person}{Mingtao Pei}, \bibinfo{person}{Yao Hu}, \bibinfo{person}{Yunde Jia}, {and} \bibinfo{person}{Qi Wu}.} \bibinfo{year}{2020}\natexlab{}.
\newblock \showarticletitle{Visual-semantic graph matching for visual grounding}. In \bibinfo{booktitle}{\emph{Proceedings of the 28th ACM International Conference on Multimedia}}. \bibinfo{pages}{4041--4050}.
\newblock


\bibitem[Kim et~al\mbox{.}(2023)]%
        {kim2023semi}
\bibfield{author}{\bibinfo{person}{Sungdong Kim}, \bibinfo{person}{Jin-Hwa Kim}, \bibinfo{person}{Jiyoung Lee}, {and} \bibinfo{person}{Minjoon Seo}.} \bibinfo{year}{2023}\natexlab{}.
\newblock \showarticletitle{Semi-parametric video-grounded text generation}.
\newblock \bibinfo{journal}{\emph{arXiv preprint arXiv:2301.11507}} (\bibinfo{year}{2023}).
\newblock


\bibitem[Koh et~al\mbox{.}(2024)]%
        {koh2024visualwebarena}
\bibfield{author}{\bibinfo{person}{Jing~Yu Koh}, \bibinfo{person}{Robert Lo}, \bibinfo{person}{Lawrence Jang}, \bibinfo{person}{Vikram Duvvur}, \bibinfo{person}{Ming~Chong Lim}, \bibinfo{person}{Po-Yu Huang}, \bibinfo{person}{Graham Neubig}, \bibinfo{person}{Shuyan Zhou}, \bibinfo{person}{Ruslan Salakhutdinov}, {and} \bibinfo{person}{Daniel Fried}.} \bibinfo{year}{2024}\natexlab{}.
\newblock \showarticletitle{Visualwebarena: Evaluating multimodal agents on realistic visual web tasks}.
\newblock \bibinfo{journal}{\emph{arXiv preprint arXiv:2401.13649}} (\bibinfo{year}{2024}).
\newblock


\bibitem[Korbar et~al\mbox{.}(2023)]%
        {korbar2023text}
\bibfield{author}{\bibinfo{person}{Bruno Korbar}, \bibinfo{person}{Yongqin Xian}, \bibinfo{person}{Alessio Tonioni}, \bibinfo{person}{Andrew Zisserman}, {and} \bibinfo{person}{Federico Tombari}.} \bibinfo{year}{2023}\natexlab{}.
\newblock \showarticletitle{Text-conditioned resampler for long form video understanding}.
\newblock \bibinfo{journal}{\emph{arXiv preprint arXiv:2312.11897}} (\bibinfo{year}{2023}).
\newblock


\bibitem[Lei et~al\mbox{.}(2021)]%
        {lei2021less}
\bibfield{author}{\bibinfo{person}{Jie Lei}, \bibinfo{person}{Linjie Li}, \bibinfo{person}{Luowei Zhou}, \bibinfo{person}{Zhe Gan}, \bibinfo{person}{Tamara~L. Berg}, \bibinfo{person}{Mohit Bansal}, {and} \bibinfo{person}{Jingjing Liu}.} \bibinfo{year}{2021}\natexlab{}.
\newblock \showarticletitle{Less is More: ClipBERT for Video-and-Language Learningvia Sparse Sampling}. In \bibinfo{booktitle}{\emph{CVPR}}.
\newblock


\bibitem[Li et~al\mbox{.}(2023)]%
        {li2023videochat}
\bibfield{author}{\bibinfo{person}{KunChang Li}, \bibinfo{person}{Yinan He}, \bibinfo{person}{Yi Wang}, \bibinfo{person}{Yizhuo Li}, \bibinfo{person}{Wenhai Wang}, \bibinfo{person}{Ping Luo}, \bibinfo{person}{Yali Wang}, \bibinfo{person}{Limin Wang}, {and} \bibinfo{person}{Yu Qiao}.} \bibinfo{year}{2023}\natexlab{}.
\newblock \bibinfo{title}{VideoChat: Chat-Centric Video Understanding}.
\newblock
\newblock
\showeprint[arxiv]{2305.06355}~[cs.CV]


\bibitem[Liu et~al\mbox{.}(2024)]%
        {liu2024visualagentbench}
\bibfield{author}{\bibinfo{person}{Xiao Liu}, \bibinfo{person}{Tianjie Zhang}, \bibinfo{person}{Yu Gu}, \bibinfo{person}{Iat~Long Iong}, \bibinfo{person}{Yifan Xu}, \bibinfo{person}{Xixuan Song}, \bibinfo{person}{Shudan Zhang}, \bibinfo{person}{Hanyu Lai}, \bibinfo{person}{Xinyi Liu}, \bibinfo{person}{Hanlin Zhao}, {et~al\mbox{.}}} \bibinfo{year}{2024}\natexlab{}.
\newblock \showarticletitle{Visualagentbench: Towards large multimodal models as visual foundation agents}.
\newblock \bibinfo{journal}{\emph{arXiv preprint arXiv:2408.06327}} (\bibinfo{year}{2024}).
\newblock


\bibitem[Liu et~al\mbox{.}(2022)]%
        {liu2022ts2}
\bibfield{author}{\bibinfo{person}{Yuqi Liu}, \bibinfo{person}{Pengfei Xiong}, \bibinfo{person}{Luhui Xu}, \bibinfo{person}{Shengming Cao}, {and} \bibinfo{person}{Qin Jin}.} \bibinfo{year}{2022}\natexlab{}.
\newblock \showarticletitle{Ts2-net: Token shift and selection transformer for text-video retrieval}. In \bibinfo{booktitle}{\emph{European conference on computer vision}}. Springer, \bibinfo{pages}{319--335}.
\newblock


\bibitem[Ma et~al\mbox{.}(2023)]%
        {ma2023vista}
\bibfield{author}{\bibinfo{person}{Fan Ma}, \bibinfo{person}{Xiaojie Jin}, \bibinfo{person}{Heng Wang}, \bibinfo{person}{Yuchen Xian}, \bibinfo{person}{Jiashi Feng}, {and} \bibinfo{person}{Yi Yang}.} \bibinfo{year}{2023}\natexlab{}.
\newblock \showarticletitle{Vista-LLaMA: Reliable Video Narrator via Equal Distance to Visual Tokens}.
\newblock \bibinfo{journal}{\emph{arXiv preprint arXiv:2312.08870}} (\bibinfo{year}{2023}).
\newblock


\bibitem[Mangalam et~al\mbox{.}(2023)]%
        {mangalam2023egoschema}
\bibfield{author}{\bibinfo{person}{Karttikeya Mangalam}, \bibinfo{person}{Raiymbek Akshulakov}, {and} \bibinfo{person}{Jitendra Malik}.} \bibinfo{year}{2023}\natexlab{}.
\newblock \showarticletitle{EgoSchema: {A} Diagnostic Benchmark for Very Long-form Video Language Understanding}. In \bibinfo{booktitle}{\emph{Advances in Neural Information Processing Systems 36: Annual Conference on Neural Information Processing Systems 2023, NeurIPS 2023, New Orleans, LA, USA, December 10 - 16, 2023}}, \bibfield{editor}{\bibinfo{person}{Alice Oh}, \bibinfo{person}{Tristan Naumann}, \bibinfo{person}{Amir Globerson}, \bibinfo{person}{Kate Saenko}, \bibinfo{person}{Moritz Hardt}, {and} \bibinfo{person}{Sergey Levine}} (Eds.).
\newblock
\urldef\tempurl%
\url{http://papers.nips.cc/paper\_files/paper/2023/hash/90ce332aff156b910b002ce4e6880dec-Abstract-Datasets\_and\_Benchmarks.html}
\showURL{%
\tempurl}


\bibitem[Mangalam et~al\mbox{.}(2024)]%
        {mangalam2024egoschema}
\bibfield{author}{\bibinfo{person}{Karttikeya Mangalam}, \bibinfo{person}{Ruslan Akshulakov}, {and} \bibinfo{person}{Jitendra Malik}.} \bibinfo{year}{2024}\natexlab{}.
\newblock \showarticletitle{Egoschema: A diagnostic benchmark for very long-form video language understanding}.
\newblock \bibinfo{journal}{\emph{Advances in Neural Information Processing Systems}}  \bibinfo{volume}{36} (\bibinfo{year}{2024}).
\newblock


\bibitem[Momeni et~al\mbox{.}(2023)]%
        {momeni2023verbs}
\bibfield{author}{\bibinfo{person}{Liliane Momeni}, \bibinfo{person}{Mathilde Caron}, \bibinfo{person}{Arsha Nagrani}, \bibinfo{person}{Andrew Zisserman}, {and} \bibinfo{person}{Cordelia Schmid}.} \bibinfo{year}{2023}\natexlab{}.
\newblock \showarticletitle{Verbs in action: Improving verb understanding in video-language models}. In \bibinfo{booktitle}{\emph{Proceedings of the IEEE/CVF International Conference on Computer Vision}}. \bibinfo{pages}{15579--15591}.
\newblock


\bibitem[Nguyen et~al\mbox{.}(2022)]%
        {nguyen2022s4nd}
\bibfield{author}{\bibinfo{person}{Eric Nguyen}, \bibinfo{person}{Karan Goel}, \bibinfo{person}{Albert Gu}, \bibinfo{person}{Gordon Downs}, \bibinfo{person}{Preey Shah}, \bibinfo{person}{Tri Dao}, \bibinfo{person}{Stephen Baccus}, {and} \bibinfo{person}{Christopher R{\'e}}.} \bibinfo{year}{2022}\natexlab{}.
\newblock \showarticletitle{S4nd: Modeling images and videos as multidimensional signals with state spaces}.
\newblock \bibinfo{journal}{\emph{Advances in neural information processing systems}}  \bibinfo{volume}{35} (\bibinfo{year}{2022}), \bibinfo{pages}{2846--2861}.
\newblock


\bibitem[OpenAI(2023)]%
        {openai2023gpt4}
\bibfield{author}{\bibinfo{person}{OpenAI}.} \bibinfo{year}{2023}\natexlab{}.
\newblock \bibinfo{title}{GPT-4 Technical Report}.
\newblock
\newblock
\showeprint[arxiv]{2303.08774}~[cs.CL]


\bibitem[Ost et~al\mbox{.}(2021)]%
        {ost2021neural}
\bibfield{author}{\bibinfo{person}{Julian Ost}, \bibinfo{person}{Fahim Mannan}, \bibinfo{person}{Nils Thuerey}, \bibinfo{person}{Julian Knodt}, {and} \bibinfo{person}{Felix Heide}.} \bibinfo{year}{2021}\natexlab{}.
\newblock \showarticletitle{Neural scene graphs for dynamic scenes}. In \bibinfo{booktitle}{\emph{Proceedings of the IEEE/CVF Conference on Computer Vision and Pattern Recognition}}. \bibinfo{pages}{2856--2865}.
\newblock


\bibitem[Papalampidi et~al\mbox{.}(2023)]%
        {papalampidi2023simple}
\bibfield{author}{\bibinfo{person}{Pinelopi Papalampidi}, \bibinfo{person}{Skanda Koppula}, \bibinfo{person}{Shreya Pathak}, \bibinfo{person}{Justin Chiu}, \bibinfo{person}{Joe Heyward}, \bibinfo{person}{Viorica Patraucean}, \bibinfo{person}{Jiajun Shen}, \bibinfo{person}{Antoine Miech}, \bibinfo{person}{Andrew Zisserman}, {and} \bibinfo{person}{Aida Nematzdeh}.} \bibinfo{year}{2023}\natexlab{}.
\newblock \showarticletitle{A Simple Recipe for Contrastively Pre-training Video-First Encoders Beyond 16 Frames}.
\newblock \bibinfo{journal}{\emph{arXiv preprint arXiv:2312.07395}} (\bibinfo{year}{2023}).
\newblock


\bibitem[Papalampidi et~al\mbox{.}(2024)]%
        {papalampidi2024simple}
\bibfield{author}{\bibinfo{person}{Pinelopi Papalampidi}, \bibinfo{person}{Skanda Koppula}, \bibinfo{person}{Shreya Pathak}, \bibinfo{person}{Justin Chiu}, \bibinfo{person}{Joe Heyward}, \bibinfo{person}{Viorica Patraucean}, \bibinfo{person}{Jiajun Shen}, \bibinfo{person}{Antoine Miech}, \bibinfo{person}{Andrew Zisserman}, {and} \bibinfo{person}{Aida Nematzdeh}.} \bibinfo{year}{2024}\natexlab{}.
\newblock \showarticletitle{A simple recipe for contrastively pre-training video-first encoders beyond 16 frames}. In \bibinfo{booktitle}{\emph{Proceedings of the IEEE/CVF Conference on Computer Vision and Pattern Recognition}}. \bibinfo{pages}{14386--14397}.
\newblock


\bibitem[Song et~al\mbox{.}(2024)]%
        {song2024moviechat}
\bibfield{author}{\bibinfo{person}{Enxin Song}, \bibinfo{person}{Wenhao Chai}, \bibinfo{person}{Guanhong Wang}, \bibinfo{person}{Yucheng Zhang}, \bibinfo{person}{Haoyang Zhou}, \bibinfo{person}{Feiyang Wu}, \bibinfo{person}{Haozhe Chi}, \bibinfo{person}{Xun Guo}, \bibinfo{person}{Tian Ye}, \bibinfo{person}{Yanting Zhang}, {et~al\mbox{.}}} \bibinfo{year}{2024}\natexlab{}.
\newblock \showarticletitle{Moviechat: From dense token to sparse memory for long video understanding}. In \bibinfo{booktitle}{\emph{Proceedings of the IEEE/CVF Conference on Computer Vision and Pattern Recognition}}. \bibinfo{pages}{18221--18232}.
\newblock


\bibitem[Sun et~al\mbox{.}(2023)]%
        {sun2023eva}
\bibfield{author}{\bibinfo{person}{Quan Sun}, \bibinfo{person}{Yuxin Fang}, \bibinfo{person}{Ledell Wu}, \bibinfo{person}{Xinlong Wang}, {and} \bibinfo{person}{Yue Cao}.} \bibinfo{year}{2023}\natexlab{}.
\newblock \showarticletitle{Eva-clip: Improved training techniques for clip at scale}.
\newblock \bibinfo{journal}{\emph{arXiv preprint arXiv:2303.15389}} (\bibinfo{year}{2023}).
\newblock


\bibitem[Sun et~al\mbox{.}(2024)]%
        {sun2024eva}
\bibfield{author}{\bibinfo{person}{Quan Sun}, \bibinfo{person}{Jinsheng Wang}, \bibinfo{person}{Qiying Yu}, \bibinfo{person}{Yufeng Cui}, \bibinfo{person}{Fan Zhang}, \bibinfo{person}{Xiaosong Zhang}, {and} \bibinfo{person}{Xinlong Wang}.} \bibinfo{year}{2024}\natexlab{}.
\newblock \showarticletitle{{EVA-CLIP-18B:} Scaling {CLIP} to 18 Billion Parameters}.
\newblock \bibinfo{journal}{\emph{CoRR}}  \bibinfo{volume}{abs/2402.04252} (\bibinfo{year}{2024}).
\newblock
\urldef\tempurl%
\url{https://doi.org/10.48550/ARXIV.2402.04252}
\showDOI{\tempurl}
\showeprint[arXiv]{2402.04252}


\bibitem[Sun et~al\mbox{.}(2022)]%
        {sun2022long}
\bibfield{author}{\bibinfo{person}{Yuchong Sun}, \bibinfo{person}{Hongwei Xue}, \bibinfo{person}{Ruihua Song}, \bibinfo{person}{Bei Liu}, \bibinfo{person}{Huan Yang}, {and} \bibinfo{person}{Jianlong Fu}.} \bibinfo{year}{2022}\natexlab{}.
\newblock \showarticletitle{Long-form video-language pre-training with multimodal temporal contrastive learning}.
\newblock \bibinfo{journal}{\emph{Advances in neural information processing systems}}  \bibinfo{volume}{35} (\bibinfo{year}{2022}), \bibinfo{pages}{38032--38045}.
\newblock


\bibitem[Sur\'is et~al\mbox{.}(2023)]%
        {suris2023vipergpt}
\bibfield{author}{\bibinfo{person}{D\'idac Sur\'is}, \bibinfo{person}{Sachit Menon}, {and} \bibinfo{person}{Carl Vondrick}.} \bibinfo{year}{2023}\natexlab{}.
\newblock \showarticletitle{ViperGPT: Visual Inference via Python Execution for Reasoning}.
\newblock \bibinfo{journal}{\emph{Proceedings of IEEE International Conference on Computer Vision (ICCV)}} (\bibinfo{year}{2023}).
\newblock


\bibitem[Wang et~al\mbox{.}(2023b)]%
        {wang2023selective}
\bibfield{author}{\bibinfo{person}{Jue Wang}, \bibinfo{person}{Wentao Zhu}, \bibinfo{person}{Pichao Wang}, \bibinfo{person}{Xiang Yu}, \bibinfo{person}{Linda Liu}, \bibinfo{person}{Mohamed Omar}, {and} \bibinfo{person}{Raffay Hamid}.} \bibinfo{year}{2023}\natexlab{b}.
\newblock \showarticletitle{Selective Structured State-Spaces for Long-Form Video Understanding}. In \bibinfo{booktitle}{\emph{{IEEE/CVF} Conference on Computer Vision and Pattern Recognition, {CVPR} 2023, Vancouver, BC, Canada, June 17-24, 2023}}. \bibinfo{publisher}{{IEEE}}, \bibinfo{pages}{6387--6397}.
\newblock
\urldef\tempurl%
\url{https://doi.org/10.1109/CVPR52729.2023.00618}
\showDOI{\tempurl}


\bibitem[Wang et~al\mbox{.}(2024)]%
        {Wang2024Vamos}
\bibfield{author}{\bibinfo{person}{Shijie Wang}, \bibinfo{person}{Qi Zhao}, \bibinfo{person}{Minh~Quan Do}, \bibinfo{person}{Nakul Agarwal}, \bibinfo{person}{Kwonjoon Lee}, {and} \bibinfo{person}{Chen Sun}.} \bibinfo{year}{2024}\natexlab{}.
\newblock \showarticletitle{Vamos: Versatile Action Models for Video Understanding}. In \bibinfo{booktitle}{\emph{Computer Vision - {ECCV} 2024 - 18th European Conference, Milan, Italy, September 29-October 4, 2024, Proceedings, Part {XII}}} \emph{(\bibinfo{series}{Lecture Notes in Computer Science}, Vol.~\bibinfo{volume}{15070})}, \bibfield{editor}{\bibinfo{person}{Ales Leonardis}, \bibinfo{person}{Elisa Ricci}, \bibinfo{person}{Stefan Roth}, \bibinfo{person}{Olga Russakovsky}, \bibinfo{person}{Torsten Sattler}, {and} \bibinfo{person}{G{\"{u}}l Varol}} (Eds.). \bibinfo{publisher}{Springer}, \bibinfo{pages}{142--160}.
\newblock
\urldef\tempurl%
\url{https://doi.org/10.1007/978-3-031-73254-6\_9}
\showDOI{\tempurl}


\bibitem[Wang et~al\mbox{.}(2025)]%
        {wang2025videoagent}
\bibfield{author}{\bibinfo{person}{Xiaohan Wang}, \bibinfo{person}{Yuhui Zhang}, \bibinfo{person}{Orr Zohar}, {and} \bibinfo{person}{Serena Yeung-Levy}.} \bibinfo{year}{2025}\natexlab{}.
\newblock \showarticletitle{Videoagent: Long-form video understanding with large language model as agent}. In \bibinfo{booktitle}{\emph{European Conference on Computer Vision}}. Springer, \bibinfo{pages}{58--76}.
\newblock


\bibitem[Wang et~al\mbox{.}(2021)]%
        {wang2021supervoxel}
\bibfield{author}{\bibinfo{person}{Yang Wang}, \bibinfo{person}{Gedas Bertasius}, \bibinfo{person}{Tae{-}Hyun Oh}, \bibinfo{person}{Abhinav Gupta}, \bibinfo{person}{Minh Hoai}, {and} \bibinfo{person}{Lorenzo Torresani}.} \bibinfo{year}{2021}\natexlab{}.
\newblock \showarticletitle{Supervoxel Attention Graphs for Long-Range Video Modeling}. In \bibinfo{booktitle}{\emph{{IEEE} Winter Conference on Applications of Computer Vision, {WACV} 2021, Waikoloa, HI, USA, January 3-8, 2021}}. \bibinfo{publisher}{{IEEE}}, \bibinfo{pages}{155--166}.
\newblock
\urldef\tempurl%
\url{https://doi.org/10.1109/WACV48630.2021.00020}
\showDOI{\tempurl}


\bibitem[Wang et~al\mbox{.}(2022)]%
        {wang2022internvideo}
\bibfield{author}{\bibinfo{person}{Yi Wang}, \bibinfo{person}{Kunchang Li}, \bibinfo{person}{Yizhuo Li}, \bibinfo{person}{Yinan He}, \bibinfo{person}{Bingkun Huang}, \bibinfo{person}{Zhiyu Zhao}, \bibinfo{person}{Hongjie Zhang}, \bibinfo{person}{Jilan Xu}, \bibinfo{person}{Yi Liu}, \bibinfo{person}{Zun Wang}, \bibinfo{person}{Sen Xing}, \bibinfo{person}{Guo Chen}, \bibinfo{person}{Junting Pan}, \bibinfo{person}{Jiashuo Yu}, \bibinfo{person}{Yali Wang}, \bibinfo{person}{Limin Wang}, {and} \bibinfo{person}{Yu Qiao}.} \bibinfo{year}{2022}\natexlab{}.
\newblock \showarticletitle{InternVideo: General Video Foundation Models via Generative and Discriminative Learning}.
\newblock \bibinfo{journal}{\emph{CoRR}}  \bibinfo{volume}{abs/2212.03191} (\bibinfo{year}{2022}).
\newblock
\urldef\tempurl%
\url{https://doi.org/10.48550/ARXIV.2212.03191}
\showDOI{\tempurl}
\showeprint[arXiv]{2212.03191}


\bibitem[Wang et~al\mbox{.}(2023a)]%
        {wang2023lifelongmemory}
\bibfield{author}{\bibinfo{person}{Yixin Wang}, \bibinfo{person}{You Yang}, {and} \bibinfo{person}{Mingzhuo Ren}.} \bibinfo{year}{2023}\natexlab{a}.
\newblock \showarticletitle{Lifelongmemory: Leveraging llms for answering queries in egocentric videos}.
\newblock \bibinfo{journal}{\emph{arXiv preprint arXiv:2312.05269}} (\bibinfo{year}{2023}).
\newblock


\bibitem[Weng et~al\mbox{.}(2025)]%
        {weng2025longvlm}
\bibfield{author}{\bibinfo{person}{Yuetian Weng}, \bibinfo{person}{Mingfei Han}, \bibinfo{person}{Haoyu He}, \bibinfo{person}{Xiaojun Chang}, {and} \bibinfo{person}{Bohan Zhuang}.} \bibinfo{year}{2025}\natexlab{}.
\newblock \showarticletitle{Longvlm: Efficient long video understanding via large language models}. In \bibinfo{booktitle}{\emph{European Conference on Computer Vision}}. Springer, \bibinfo{pages}{453--470}.
\newblock


\bibitem[Woo et~al\mbox{.}(2018)]%
        {woo2018linknet}
\bibfield{author}{\bibinfo{person}{Sanghyun Woo}, \bibinfo{person}{Dahun Kim}, \bibinfo{person}{Donghyeon Cho}, {and} \bibinfo{person}{In~So Kweon}.} \bibinfo{year}{2018}\natexlab{}.
\newblock \showarticletitle{Linknet: Relational embedding for scene graph}.
\newblock \bibinfo{journal}{\emph{Advances in neural information processing systems}}  \bibinfo{volume}{31} (\bibinfo{year}{2018}).
\newblock


\bibitem[Wu et~al\mbox{.}(2022)]%
        {wu2022memvit}
\bibfield{author}{\bibinfo{person}{Chao{-}Yuan Wu}, \bibinfo{person}{Yanghao Li}, \bibinfo{person}{Karttikeya Mangalam}, \bibinfo{person}{Haoqi Fan}, \bibinfo{person}{Bo Xiong}, \bibinfo{person}{Jitendra Malik}, {and} \bibinfo{person}{Christoph Feichtenhofer}.} \bibinfo{year}{2022}\natexlab{}.
\newblock \showarticletitle{MeMViT: Memory-Augmented Multiscale Vision Transformer for Efficient Long-Term Video Recognition}. In \bibinfo{booktitle}{\emph{{IEEE/CVF} Conference on Computer Vision and Pattern Recognition, {CVPR} 2022, New Orleans, LA, USA, June 18-24, 2022}}. \bibinfo{publisher}{{IEEE}}, \bibinfo{pages}{13577--13587}.
\newblock
\urldef\tempurl%
\url{https://doi.org/10.1109/CVPR52688.2022.01322}
\showDOI{\tempurl}


\bibitem[Xiao et~al\mbox{.}(2021)]%
        {xiao2021next}
\bibfield{author}{\bibinfo{person}{Junbin Xiao}, \bibinfo{person}{Xindi Shang}, \bibinfo{person}{Angela Yao}, {and} \bibinfo{person}{Tat-Seng Chua}.} \bibinfo{year}{2021}\natexlab{}.
\newblock \showarticletitle{Next-qa: Next phase of question-answering to explaining temporal actions}. In \bibinfo{booktitle}{\emph{Proceedings of the IEEE/CVF conference on computer vision and pattern recognition}}. \bibinfo{pages}{9777--9786}.
\newblock


\bibitem[Xiao et~al\mbox{.}(2023)]%
        {xiaovgt}
\bibfield{author}{\bibinfo{person}{J. Xiao}, \bibinfo{person}{P. Zhou}, \bibinfo{person}{A. Yao}, \bibinfo{person}{Y. Li}, \bibinfo{person}{R. Hong}, \bibinfo{person}{S. Yan}, {and} \bibinfo{person}{T. Chua}.} \bibinfo{year}{2023}\natexlab{}.
\newblock \showarticletitle{Contrastive Video Question Answering via Video Graph Transformer}.
\newblock \bibinfo{journal}{\emph{IEEE Transactions on Pattern Analysis; Machine Intelligence}} \bibinfo{volume}{45}, \bibinfo{number}{11} (\bibinfo{date}{nov} \bibinfo{year}{2023}), \bibinfo{pages}{13265--13280}.
\newblock
\showISSN{1939-3539}


\bibitem[Xu et~al\mbox{.}(2023)]%
        {xu2023retrievalbased}
\bibfield{author}{\bibinfo{person}{Jiaqi Xu}, \bibinfo{person}{Cuiling Lan}, \bibinfo{person}{Wenxuan Xie}, \bibinfo{person}{Xuejin Chen}, {and} \bibinfo{person}{Yan Lu}.} \bibinfo{year}{2023}\natexlab{}.
\newblock \bibinfo{title}{Retrieval-based Video Language Model for Efficient Long Video Question Answering}.
\newblock
\newblock
\showeprint[arxiv]{2312.04931}~[cs.CV]


\bibitem[Yang et~al\mbox{.}(2021)]%
        {yang2021just}
\bibfield{author}{\bibinfo{person}{Antoine Yang}, \bibinfo{person}{Antoine Miech}, \bibinfo{person}{Josef Sivic}, \bibinfo{person}{Ivan Laptev}, {and} \bibinfo{person}{Cordelia Schmid}.} \bibinfo{year}{2021}\natexlab{}.
\newblock \showarticletitle{Just ask: Learning to answer questions from millions of narrated videos}. In \bibinfo{booktitle}{\emph{Proceedings of the IEEE/CVF international conference on computer vision}}. \bibinfo{pages}{1686--1697}.
\newblock


\bibitem[Yang et~al\mbox{.}(2022b)]%
        {yang2022frozenbilm}
\bibfield{author}{\bibinfo{person}{Antoine Yang}, \bibinfo{person}{Antoine Miech}, \bibinfo{person}{Josef Sivic}, \bibinfo{person}{Ivan Laptev}, {and} \bibinfo{person}{Cordelia Schmid}.} \bibinfo{year}{2022}\natexlab{b}.
\newblock \showarticletitle{Zero-Shot Video Question Answering via Frozen Bidirectional Language Models}. In \bibinfo{booktitle}{\emph{Advances in Neural Information Processing Systems 35: Annual Conference on Neural Information Processing Systems 2022, NeurIPS 2022, New Orleans, LA, USA, November 28 - December 9, 2022}}, \bibfield{editor}{\bibinfo{person}{Sanmi Koyejo}, \bibinfo{person}{S.~Mohamed}, \bibinfo{person}{A.~Agarwal}, \bibinfo{person}{Danielle Belgrave}, \bibinfo{person}{K.~Cho}, {and} \bibinfo{person}{A.~Oh}} (Eds.).
\newblock
\urldef\tempurl%
\url{http://papers.nips.cc/paper\_files/paper/2022/hash/00d1f03b87a401b1c7957e0cc785d0bc-Abstract-Conference.html}
\showURL{%
\tempurl}


\bibitem[Yang et~al\mbox{.}(2022a)]%
        {yang2022panoptic}
\bibfield{author}{\bibinfo{person}{Jingkang Yang}, \bibinfo{person}{Yi~Zhe Ang}, \bibinfo{person}{Zujin Guo}, \bibinfo{person}{Kaiyang Zhou}, \bibinfo{person}{Wayne Zhang}, {and} \bibinfo{person}{Ziwei Liu}.} \bibinfo{year}{2022}\natexlab{a}.
\newblock \showarticletitle{Panoptic scene graph generation}. In \bibinfo{booktitle}{\emph{European Conference on Computer Vision}}. Springer, \bibinfo{pages}{178--196}.
\newblock


\bibitem[Yang et~al\mbox{.}(2020)]%
        {yang2020gives}
\bibfield{author}{\bibinfo{person}{Jianing Yang}, \bibinfo{person}{Yuying Zhu}, \bibinfo{person}{Yongxin Wang}, \bibinfo{person}{Ruitao Yi}, \bibinfo{person}{Amir Zadeh}, {and} \bibinfo{person}{Louis-Philippe Morency}.} \bibinfo{year}{2020}\natexlab{}.
\newblock \showarticletitle{What gives the answer away? question answering bias analysis on video qa datasets}.
\newblock \bibinfo{journal}{\emph{arXiv preprint arXiv:2007.03626}} (\bibinfo{year}{2020}).
\newblock


\bibitem[Ye et~al\mbox{.}(2023)]%
        {ye2023hitea}
\bibfield{author}{\bibinfo{person}{Qinghao Ye}, \bibinfo{person}{Guohai Xu}, \bibinfo{person}{Ming Yan}, \bibinfo{person}{Haiyang Xu}, \bibinfo{person}{Qi Qian}, \bibinfo{person}{Ji Zhang}, {and} \bibinfo{person}{Fei Huang}.} \bibinfo{year}{2023}\natexlab{}.
\newblock \showarticletitle{Hitea: Hierarchical temporal-aware video-language pre-training}. In \bibinfo{booktitle}{\emph{Proceedings of the IEEE/CVF International Conference on Computer Vision}}. \bibinfo{pages}{15405--15416}.
\newblock


\bibitem[Yu et~al\mbox{.}(2023)]%
        {yu2023self}
\bibfield{author}{\bibinfo{person}{Shoubin Yu}, \bibinfo{person}{Jaemin Cho}, \bibinfo{person}{Prateek Yadav}, {and} \bibinfo{person}{Mohit Bansal}.} \bibinfo{year}{2023}\natexlab{}.
\newblock \showarticletitle{Self-Chained Image-Language Model for Video Localization and Question Answering}.
\newblock \bibinfo{journal}{\emph{NeurIPS}} (\bibinfo{year}{2023}).
\newblock


\bibitem[Yu et~al\mbox{.}(2024)]%
        {sevila}
\bibfield{author}{\bibinfo{person}{Sangho Yu}, \bibinfo{person}{Jaemin Cho}, \bibinfo{person}{Prateek Yadav}, {and} \bibinfo{person}{Mohit Bansal}.} \bibinfo{year}{2024}\natexlab{}.
\newblock \showarticletitle{Self-chained image-language model for video localization and question answering}.
\newblock \bibinfo{journal}{\emph{Advances in Neural Information Processing Systems}}  \bibinfo{volume}{36} (\bibinfo{year}{2024}).
\newblock


\bibitem[Zellers et~al\mbox{.}(2018)]%
        {zellers2018neural}
\bibfield{author}{\bibinfo{person}{Rowan Zellers}, \bibinfo{person}{Mark Yatskar}, \bibinfo{person}{Sam Thomson}, {and} \bibinfo{person}{Yejin Choi}.} \bibinfo{year}{2018}\natexlab{}.
\newblock \showarticletitle{Neural motifs: Scene graph parsing with global context}. In \bibinfo{booktitle}{\emph{Proceedings of the IEEE conference on computer vision and pattern recognition}}. \bibinfo{pages}{5831--5840}.
\newblock


\bibitem[Zhang et~al\mbox{.}(2023a)]%
        {zhang2023simple}
\bibfield{author}{\bibinfo{person}{Ce Zhang}, \bibinfo{person}{Taixi Lu}, \bibinfo{person}{Md~Mohaiminul Islam}, \bibinfo{person}{Ziyang Wang}, \bibinfo{person}{Shoubin Yu}, \bibinfo{person}{Mohit Bansal}, {and} \bibinfo{person}{Gedas Bertasius}.} \bibinfo{year}{2023}\natexlab{a}.
\newblock \showarticletitle{A simple llm framework for long-range video question-answering}.
\newblock \bibinfo{journal}{\emph{arXiv preprint arXiv:2312.17235}} (\bibinfo{year}{2023}).
\newblock


\bibitem[Zhang et~al\mbox{.}(2023b)]%
        {llovi}
\bibfield{author}{\bibinfo{person}{Chunting Zhang}, \bibinfo{person}{Thomas Lu}, \bibinfo{person}{Md~Mohaiminul Islam}, \bibinfo{person}{Zichen Wang}, \bibinfo{person}{Sangho Yu}, \bibinfo{person}{Mohit Bansal}, {and} \bibinfo{person}{Gedas Bertasius}.} \bibinfo{year}{2023}\natexlab{b}.
\newblock \showarticletitle{A simple llm framework for long-range video question-answering}.
\newblock \bibinfo{journal}{\emph{arXiv preprint arXiv:2312.17235}} (\bibinfo{year}{2023}).
\newblock


\bibitem[Zhang et~al\mbox{.}(2024)]%
        {zhang2024flash}
\bibfield{author}{\bibinfo{person}{Haoji Zhang}, \bibinfo{person}{Yiqin Wang}, \bibinfo{person}{Yansong Tang}, \bibinfo{person}{Yong Liu}, \bibinfo{person}{Jiashi Feng}, \bibinfo{person}{Jifeng Dai}, {and} \bibinfo{person}{Xiaojie Jin}.} \bibinfo{year}{2024}\natexlab{}.
\newblock \showarticletitle{Flash-VStream: Memory-Based Real-Time Understanding for Long Video Streams}.
\newblock \bibinfo{journal}{\emph{arXiv preprint arXiv:2406.08085}} (\bibinfo{year}{2024}).
\newblock


\bibitem[Zhao et~al\mbox{.}(2023)]%
        {lavila}
\bibfield{author}{\bibinfo{person}{Yue Zhao}, \bibinfo{person}{Ishan Misra}, \bibinfo{person}{Philipp Krähenbühl}, {and} \bibinfo{person}{Rohit Girdhar}.} \bibinfo{year}{2023}\natexlab{}.
\newblock \showarticletitle{Learning video representations from large language models}. In \bibinfo{booktitle}{\emph{CVPR}}.
\newblock


\end{thebibliography}

\end{document}